\documentclass{aa}

\usepackage[colorlinks=true,linkcolor=blue,citecolor    = blue]{hyperref}
\usepackage{graphicx}
\usepackage{txfonts}
\usepackage{float}
\usepackage{rotating,tabularx}
\usepackage{multirow}
\usepackage{afterpage}
\usepackage{bigstrut}
\usepackage[table,xcdraw]{xcolor}
\usepackage[flushleft]{threeparttable }
\usepackage{amsmath}
\usepackage{enumitem}
\usepackage{amsfonts,color}
\usepackage{amssymb,float}
\usepackage[utf8]{inputenc}
\usepackage{color}
\usepackage{etoolbox}
\usepackage{appendix}
\usepackage{url}
\usepackage{dirtytalk}
\usepackage{float}
\usepackage{fontawesome}

\newcommand{\ra}[1]{\renewcommand{\arraystretch}{#1}}
\usepackage{booktabs}
\usepackage[graphicx]{realboxes}
\usepackage{adjustbox}
\usepackage{multirow}
\usepackage{rotating,tabularx}
\usepackage{afterpage}
\usepackage{tablefootnote}

\usepackage{tikz}

\usepackage{bigstrut}

\definecolor{lime}{HTML}{A6CE39}
\DeclareRobustCommand{\orcidicon}{%
    \begin{tikzpicture}
    \draw[lime, fill=lime] (0,0) 
    circle [radius=0.16] 
    node[white] {{\fontfamily{qag}\selectfont \tiny ID}};
    \draw[white, fill=white] (-0.0625,0.095) 
    circle [radius=0.007];
    \end{tikzpicture}
    \hspace{-2mm}
}

\foreach \x in {A, ..., Z}{%
    \expandafter\xdef\csname orcid\x\endcsname{\noexpand\href{https://orcid.org/\csname orcidauthor\x\endcsname}{\noexpand\orcidicon}}
}
\newcommand{\orcid}[1]{\href{https://orcid.org/#1}{\textcolor[HTML]{A6CE39}{\orcidicon}}}

\newcommand{\gaia}{\textit{Gaia}}
\newcommand{\hst}{\textit{HST}}
\newcommand{\Ho}{$H_0$}
\newcommand{\kms}{$\mathrm{km\,s^{-1}}$}

\begin{document}

   \title{A 0.9\% calibration of the Galactic Cepheid luminosity scale \\
   based on {\it Gaia} DR3 data of open clusters and Cepheids \thanks{Tables \ref{tab:RV} - \ref{tab:Copper}, \ref{tab:GaiaSample}, \ref{tab:metal} \ and \ref{tab:GaiaSample2} are  available in electronic form
at the CDS via anonymous ftp to cdsarc.cds.unistra.fr (130.79.128.5)
or via https://cdsarc.cds.unistra.fr/cgi-bin/qcat?J/A+A/}   }
   \titlerunning{Cepheid luminosity calibration based on clusters and {\it Gaia} DR3} 
    \authorrunning{M.~Cruz Reyes \& R.I.~Anderson}

   \author{Mauricio Cruz Reyes \inst{1}\orcid{0000-0003-2443-173X}
          \and Richard I. Anderson\inst{1}\orcid{0000-0001-8089-4419}} 
   \institute{Institute of Physics, Laboratory of Astrophysics, \'Ecole Polytechnique F\'ed\'erale de Lausanne (EPFL), 1290 Versoix, Switzerland 
    \email{mauricio.cruzreyes@epfl.ch, richard.anderson@epfl.ch}   }
   \date{Received 19 August 2022; accepted 27 January 2023}

  \abstract
  {
  We have conducted a search for open clusters in the vicinity of classical Galactic Cepheids based on high-quality astrometry from the third data release (DR3) of the ESA mission \textit{Gaia} to improve the calibration of the Leavitt law (LL). Our approach requires no prior knowledge of existing clusters, allowing us to both detect new host clusters and cross-check previously reported associations. Our Gold sample consists of 34 Cepheids residing in 28 open clusters, including 27 fundamental mode and seven overtone Cepheids. Three new bona fide cluster Cepheids are reported (V0378 Cen, ST Tau, and GH Lup) and the host cluster identifications for three others (VW Cru, IQ Nor, and SX Vel) are corrected. The fraction of Cepheids occurring in open clusters within $2\,$kpc of the Sun is $f_{\mathrm{CC,2kpc}} = 0.088^{+0.029}_{-0.019}$.  Nonvariable cluster members allow us to determine cluster parallaxes to $\sim 7\,\mu$as in the range $12.5 < G < 17$\,mag, where recent studies found that parallax corrections by \cite{lindegren2021gaia2} (L21) are accurate and require no further offset corrections. By comparing Cepheids in MW clusters to Cepheids in the LMC, we confirm these independent results and the adequacy of the L21 corrections for the cluster members in this range. 
  By combining cluster and field Cepheids, we calibrate the LL for several individual photometric passbands, together with reddening-free Wesenheit magnitudes based on {\it Gaia} and \hst\ photometry, while solving for the residual offset applicable to Cepheid parallaxes, $\Delta \varpi_{\mathrm{Cep}}$.  The most direct comparison of our results with the SH0ES distance ladder yields excellent ($0.3\sigma$) agreement for both the absolute magnitude of a $10$\,d solar metallicity Cepheid in the near-IR \hst\ Wesenheit magnitudes, $M_{H,1}^W=-5.914\pm0.017$\,mag, and the residual parallax offset, $\Delta \varpi_{\mathrm{Cep}}=-13\pm5\,\mu$as. Despite the use of a common set of photometry, this is an important cross-check of the recent Hubble constant measurement by \cite{2022arXiv220801045R} based on independently determined cluster membership and average parallaxes.  Using the larger sample of $26$ Gold cluster Cepheids and $225$ MW Cepheids with recent \gaia\ DR3 astrometry and photometry, we determine $M_{G,1}^W = -6.051\pm 0.020$\,mag 
  in the optical \gaia\ Wesenheit magnitude at the sample average iron abundance of ($\langle \mathrm{[Fe/H]} \rangle = 0.069$)  and $\Delta \varpi_{\mathrm{Cep}}=-22 \pm 3\,\mu$as. Correcting to solar metallicity yields $M_{G,1}^W = -6.004 \pm 0.019$\,mag and $\Delta \varpi_{\mathrm{Cep}}=-19 \pm 3\,\mu$as. 
  These results mark the currently most accurate absolute calibrations of the Cepheid luminosity scale based purely on observations of Milky Way Cepheids, and it is also the most precise determination of the residual Cepheid parallax offset at a significance of $6-7\sigma$.}


   \maketitle

\section{Introduction}
 
The absolute calibration of the classical Cepheid
luminosity scale is fundamental for a distance estimation in the nearby Universe and the accurate measurement of Hubble's constant, \Ho. The third data release (DR3) of the ESA mission \textit{Gaia} has provided astrometry of unprecedented quantity and quality \citep{prusti2016gaia,brown2021gaia} for approximately 1.5 billion stars in the magnitude range $3 < G < 21$, including  14992 classical Cepheid stars \citep{gaiadr3.vari,gaiadr3.cepheid} with an average parallax uncertainty of $70\,\mu$as. Because the parallax is generally considered the gold standard of geometric distance measurements, the \gaia\ parallaxes are of crucial importance for the absolute calibration of Leavitt's law \citet[henceforth: LL]{Leavitt1912}, also known as the period-luminosity relation, and they are of great interest for all further applications of Cepheids as distance tracers. In particular, \gaia\  parallaxes are required to clarify the implications of the current $5\sigma$ discrepancy between the value of \Ho\ measured using a distance ladder composed of classical Cepheids and type Ia supernovae \citep[e.g.,][]{riess2021comprehensive} and the value of \Ho\ inferred from observations of the ESA mission {\it Planck}  of the cosmic microwave background assuming a flat $\Lambda$CDM Universe \citep{Planck2018-VI}. 

However, \gaia-based LL calibrations based on Cepheid parallaxes must currently simultaneously solve for a residual parallax offset due to systematics of the \gaia\  data processing in addition to the LL intercept and slope  \citep[e.g.,][]{riess2021cosmic}. Because this simultaneous parallax offset determination reduces the precision to which \gaia\ can calibrate the LL, strategies for mitigating this problem are needed.

\citet[henceforth: L21]{lindegren2021gaia2} derived corrections to the zeropoint offset of about $10-30\,\mu$arcsec, whose exact value depends nontrivially on the magnitude of the observed source, its position in the sky, and its color. Several studies (not necessarily based on Cepheids) have investigated residual (compared to Lindegren's correction) zeropoint offsets, generally finding good agreement at magnitudes ($G \gtrsim 13$\,mag) at which L21 is well calibrated \citep[e.g.,][]{huang2021parallax, riess2021cosmic,el2021million}, whereas an offset remains at the brighter end, where the L21 calibration was based on fewer sources.  The origins of these residual offsets are complex and not yet fully understood, although it is likely that they originate from differences between the Cepheid and quasar samples, with Cepheids being systematically brighter, of redder intrinsic color, and photometrically  and chromatically variable. Moreover, the Milky Way Cepheids that were used to calibrate the LL fall within a magnitude range ($G \lesssim 13$\,mag) that requires special observational and data-processing steps to avoid saturation (including the gating mechanism to avoid saturation and changing from 2D to 1D image processing for the astrometric model, cf. L21).

An interesting possibility for avoiding difficulties related to this zeropoint systematic could be the use of parallax information derived from stars that are observationally as similar as possible to the objects used to determine the \gaia\  systematics. Because Cepheids are relatively young stars ($< 300$\,Myr), they are occasionally found in open star clusters \citep[cf.][and references therein]{anderson2013cepheids}, whose brightest main-sequence members will tend to be bluer than Cepheids, and several magnitudes fainter. At the same time, open clusters contain many stars, so that an average cluster parallax will benefit from a $\sqrt{N}$ improvement in precision, eventually limited by the angular covariance of the \gaia\  parallaxes \citep{lindegren2021gaia2,apellaniz2021validation,vasiliev2021gaia,zinn2021}.

The currently most common approach to identifying cluster Cepheids is to consider cluster input lists from studies based on \gaia\ astrometry \citep{cantat2020clusters,hunt2021improving,castro2021hunting,zhou2021galactic,he2022new} and to then compare the astrometric parameters of Cepheids with the average cluster parameters \citep{anderson2013cepheids,breuval2020milky,zhou2021galactic,medina2021revisited}. However, there is no guarantee that all Cepheid-hosting clusters have been detected so far, and the selection function of clusters is not well known. It is also rather common for Cepheids to reside in the coronae of their host clusters, that is, farther from the center than the typical cluster core radius of $\sim 4\,$pc \citep[e.g.,][]{anderson2013cepheids}. This is to some extent expected from the clustered star formation process that causes the majority of birth clusters to disperse into the field over timescales of tens of million years \citep{dinnbier2022dynamical}. Tidal deformations further cause cluster shapes to deviate from circular over hundreds of million years, thus breaking the symmetry of the appearance and complicating the detection of cluster members against a highly contaminated background \citep{Boffin2022Tidal}.  Additionally, it is quite common for multiple clusters to exist relatively close to each other on the sky \citep{turner1998search} because of the high density of clusters in spiral arms and the superposition on sky of multiple spiral arms. Substantial and spatially variable extinction can further complicate the issue. 
To most reliably determine the most complete sample of cluster Cepheids detectable with \gaia\ DR3 data, we therefore adopted the approach of searching for clusters in the vicinity of Cepheids, rather than the other way around. 

A major improvement of the extragalactic distance ladder built by the \emph{SH0ES} project \citep{riess2021comprehensive} has been the photometric homogeneity of Cepheid observations carried out exclusively in the Hubble Space Telescope (\textit{HST}) photometric system. With the release of time-series observations in \gaia\  DR3, there is now an additional  data set of very high quality, well-resolved multichromatic observations based on a well-characterized and homogeneous photometric system that includes observations of Milky Way and Local Group Cepheids, reaching Cepheids as far as M31 and M33  \citep{gaiadr3.gaps}, albeit with increased uncertainties due to higher instrumental noise and higher crowding. The goal of this paper is to leverage these unprecedented data sets to achieve the most accurate absolute calibration of the MW LL in well-characterized filters, notably including the reddening-free near-IR \hst\ Wesenheit function used by the SH0ES team to measure the Hubble constant \citep{riess2021comprehensive}, while simultaneously solving for the residual parallax offset of Cepheids.

This article is organized as follows. Section \ref{sec:Method} describes our method for detecting and estimating the parameters of clusters in the physical vicinity of MW Cepheids based on \gaia\ data and the estimation of membership probabilities for the Cepheids. Section \ref{sec:CC} {separates the sample of cluster Cepheid candidates into Gold, Silver, and Bronze samples.  Section \ref{sec:PLR} presents the simultaneous calibration of the Cepheid LL in multiple photometric bands and an LMC-based cross-check of the L21 corrections applied to cluster member stars. Section \ref{sec:Discussion} presents an additional discussion, and Sect.~\ref{sec:Conclusions} lists our conclusions. Additional tables and figures are provided in the online appendix.

\section{Method}\label{sec:Method}

The starting point of our analysis was the list of positions of 3352 Milky Way classical Cepheids that were classified by the OGLE collaboration based on a large combination of all-sky time-series survey data \citep{pietrukowicz2021classical}, which we extended by 230 additional classical Cepheids that were reported by \gaia\ DR3 in June 2022 \citep{gaiadr3.cepheid,gaiadr3.vari}. While there can be disagreements over Cepheid classifications, especially for overtone Cepheids with sinusoidal light curves, we note that the list by \citet{pietrukowicz2021classical} was used to validate the \gaia\ DR3 sample, and that the  samples of MW Cepheids overlap by $\sim 85\%$ . Extragalactic Cepheids and Cepheids that are too distant for identifying clusters were removed from the \gaia\ DR3 sample by the quality cuts explained in Sec.\,\ref{sec:clustering} and by requiring Cepheids to be brighter than $G < 16$\,mag.   For each Cepheid we considered, we retrieved all stars within a radius of one degree from the \gaia\ archive\footnote{\url{https://gea.esac.esa.int/archive/}}  and then searched for host clusters as explained in the following and as illustrated schematically in Fig.\,\ref{fig:flow}.

\begin{figure}
    \centering
    \includegraphics{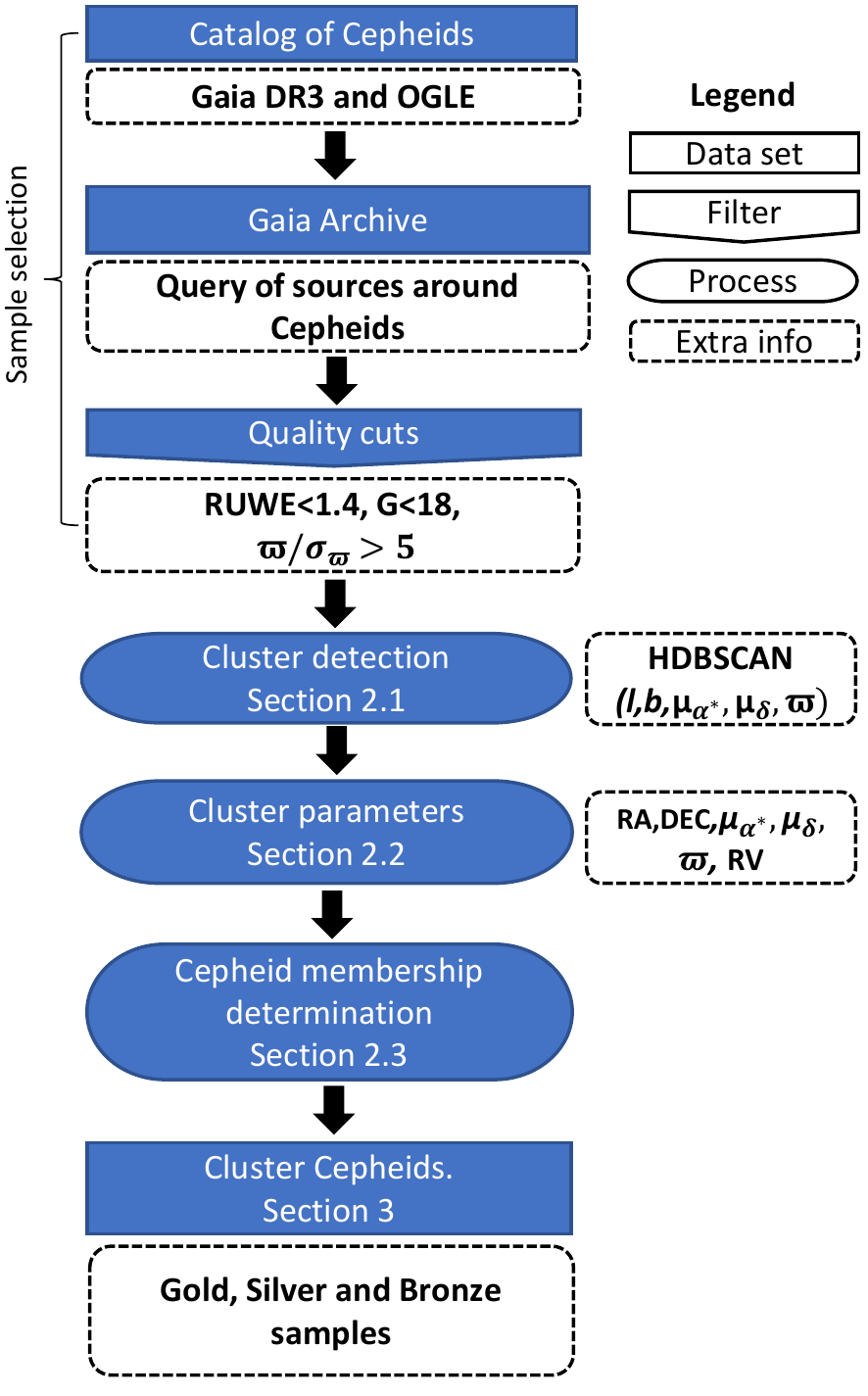}
    \caption{Schematic overview of the pipeline designed to detect cluster Cepheids. }
    \label{fig:flow}
\end{figure}

\subsection{Cluster detection} \label{sec:clustering}
Because clusters are gravitationally bound systems, cluster members share similar positions  (RA, DEC), proper motions ($\mu_{\alpha^{*}},\mu_{\delta}$), parallaxes ($\varpi$), and radial velocities. Thus, stars belonging to a common cluster can be separated from fore- or background stars as overdensities in the multidimensional space spanned by the available membership constraints.

\gaia\ DR3 provides information for all six of these parameters, although radial velocity information is only available for a rather limited number of stars owing to the faintness of most member stars. Hence, our analysis employs only positions, proper motions, and parallaxes for the cluster identification. Where available, radial velocity information was used to assess membership probabilities of Cepheids (cf. Sec.\,\ref{sec:likelihood}).

We detected clusters using the publicly available code  called hierarchical density-based spatial clustering of applications with noise \citep[HDBSCAN]{McInnes2017}. As is common practice, we included only stars in our analyis whose parallax signal-to-noise ratio $\varpi/\sigma_\varpi \geq 5$, whose renormalised unit weight error (RUWE) is smaller than $1.4$ to exclude sources with poor astrometry \citep[e.g., due to the presence of companions]{fabricius2021gaia}, and that are brighter than  $G = 18\,$mag (parameter \texttt{phot\_g\_mean\_mag} in table \texttt{gaiadr3.gaia\_source}), where the \gaia\ astrometry is most precise. In practice, this magnitude cut represents no serious limitation for our work and allows us to clearly recover the main sequences of Cepheid-hosting clusters that are several magnitudes fainter than their Cepheid members. \gaia\ parallaxes of all stars considered for membership were corrected for systematics using the recipe provided by L21.

The code HDBSCAN uses the $n-$dimensional distance between objects to identify overdense regions. As Cepheids and open clusters are located in the Galactic plane, we used Galactic coordinates ($l, b$) for the positional constraints rather than RA and DEC. The ability of HDBSCAN to detect arbitrarily shaped clusters was particularly useful for our purposes because the physical shape of  clusters in various stages of dispersal was not known a priori. The only fixed input parameter required by HDBSCAN is the number of stars $p$ that are expected to qualify an overdensity as a cluster. Deviations of the number of cluster stars $s$ from $p$ will cause overdensities with $s < p$ to remain undetected by HDBSCAN and may sometimes result in a single cluster being split into multiple parts if $s > p$. To ensure that our analysis was not  sensitive to these undesirable side effects, we repeated our analysis using ten different values for $p$ ranging from 10 to 100 in increments 10 and found a consistent number of cluster members in each case. The mean (median) number of member stars reported per cluster is 230 (152) (cf. Sect.\, \ref{sec:CC}). 

Following \citet{castro2018new} and \citet{hunt2021improving},  we rescaled each of the \gaia\  astrometric parameters to variables with zero mean and unit standard deviation  by subtracting the mean from each parameter  and rescaling parameters such that the $25-75\%$ percentile has unit variance. This procedure ensures equal weighting among the five dimensions and improves robustness against outliers.

Inspection of the parallax distributions returned by HDBSCAN revealed  outliers in parallax. To retain only likely cluster members, we  determined the mode of the parallax distribution returned by HDBSCAN and retained all cluster members whose parallaxes agreed to within $3$ standard deviations of a Gaussian fit to the parallax distribution centered on the mode.

At distances beyond $2$\,kpc, cluster identification becomes increasingly limited due to the current parallax and proper motion uncertainties of \textit{Gaia}.  
Because our goal of calibrating the Galactic LL requires utmost accuracy and precision, we prioritized greater purity (lower contamination) at the potential cost of completeness. We thus visually inspected all identified cluster candidates to ensure that cluster stars were overdense in each of the membership constraints considered and that the resulting color-magnitude diagrams indicated a coeval population being detected, as evidenced by a clearly visible main sequence. Additionally, we discarded clusters in which a majority of main-sequence stars exceeded the brightness of their candidate Cepheid members.

For each cluster, HDBSCAN provided a list of likely cluster members together with membership probabilities. By design, all identified clusters were within the projected vicinity of Cepheids. However, these same Cepheids were not necessarily selected as cluster members by HDBSCAN, requiring a separate membership analysis for Cepheids in the detected clusters (cf. Sec.\,\ref{sec:likelihood}).

\subsection{Cluster parameters\label{sec:ClParams}}

For each cluster that passed the first visual screening, we computed the center position in RA and DEC. We additionally computed  averages and dispersions in both proper motion directions, parallax, and radial velocity, where available. 
 
\subsubsection{Cluster parallaxes\label{sec:ClPlx}}

\begin{figure*} 
    \centering
    \includegraphics[width=0.99\textwidth]{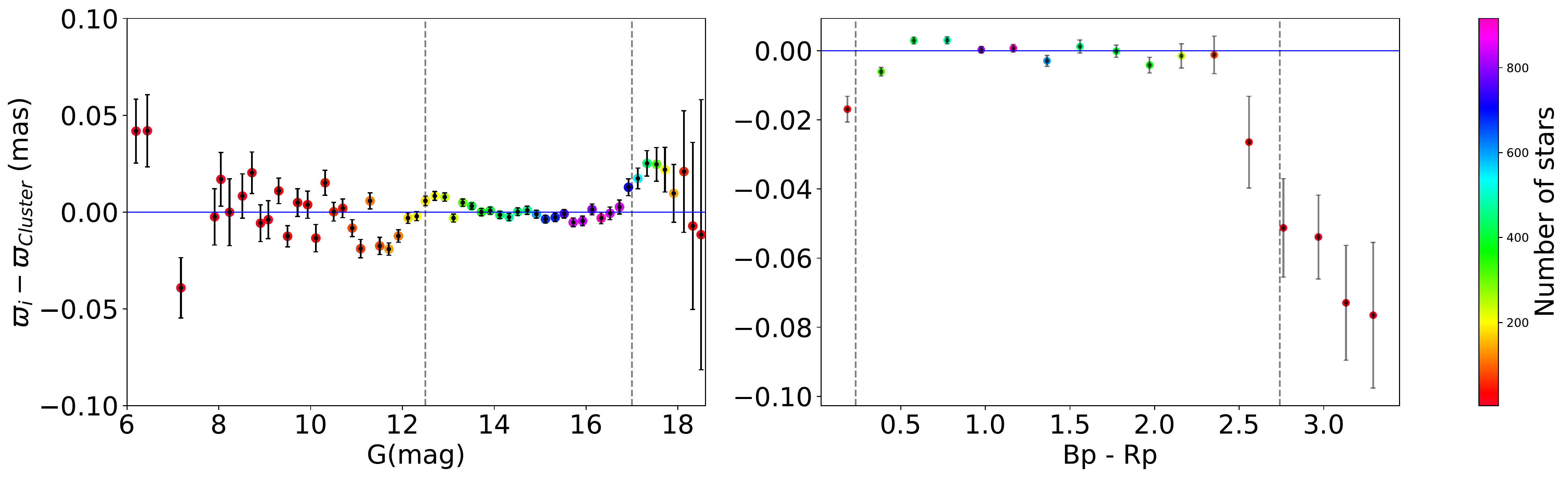}
    \caption{Difference between individual and cluster average parallax for all member stars considered.    \textit{Left:} Comparison of the parallax difference as a function of the G magnitude, where the number of stars per bin is color-coded according to the color bar on the right. \textit{Right:} Same as the left plot, but as a function of the color $Bp-Rp$.  The vertical dotted lines in both panels illustrate the magnitude and color range we used to estimate the cluster parallaxes.       }
    \label{fig:dispersion_gmag}
\end{figure*}

For a given source, the \gaia\  parallax systematics are well known to depend   on its sky position as well as its  magnitude and color (L21). Magnitude and color trends are likely related to the sophisticated on-board processing of \textit{Gaia,} which was implemented to avoid saturation across the extreme dynamic range of the survey (limit $21.7$\,mag). Using the cluster members returned by HDBSCAN, we investigated whether an optimal magnitude and color range could be identified to obtain the most reliable and precise average cluster parallax. We calculated the deviation from the cluster average, $ \Delta \varpi = \langle \varpi \rangle - \varpi_{i}$, for all member stars of all host clusters. We combined all values of $ \Delta \varpi$ into a single set,  which we divided into bins of $0.2$\,mag in $G  $ band. For each bin, we estimated the weighted mean and the weighted error of $ \Delta \varpi $. Figure~\ref{fig:dispersion_gmag} illustrates this result and shows a noticeable decrease in the variance of $\Delta \varpi$ for $12.5 < G <17$ and $0.23 <Bp - Rp< 2.75$. The observed variance is dominated by the uncertainty in parallaxes for fainter stars and by lower statistics for   bright stars ($G \lesssim 9$\,mag). Systematic trends at $G < 12.5$\,mag can be partially due to the gating mechanism of \textit{Gaia} or to differences in photometric processing \footnote{According to Fig.\,1 in L21, no gating is applied to stars fainter than $12.5$\,mag. However, the WC0b and WC1 calibration models of the astrometric field overlap in the range $12.5 < G < 13$\,mag, which implies a transition from 2D images to binned 1D images, respectively.}. 
We note that the exact magnitude range is not critical for the estimation of the cluster parallaxes. For example, restricting the magnitude range further to $13.5-17$\,mag changes the mean cluster parallax by less than $ 2\,\mu$as, while increasing the uncertainty in average parallax for clusters with $\lesssim 100$ members (e.g., CWNU 175 or vdBergh 1) by approximately $1\,\mu$as (cf. Sect.\,\ref{sec:CC}). Because of these clear and consistent trends and to avoid sensitivity to gating-related issues, we adopted the range of $12.5<G<17$\,mag as the optimal range for determining high-fidelity average cluster parallaxes and their uncertainties. We further restricted the color range of member stars to  $0.23 <Bp - Rp< 2.75$ to avoid the color range for which Fig.~\ref{fig:dispersion_gmag} shows increasing deviations from zero residuals, accompanied by increasing uncertainties due to low statistics.

Several studies have shown the existence of nonzero residual parallax offsets for stars brighter than $G < 13$\,mag after the L21 corrections were applied \citep[e.g.,][]{huang2021parallax,zinn2021validation,el2021million,riess2021cosmic,2022arXiv220801045R}. However, analyses of open and globular clusters, as well as of the LMC, have shown the L21 procedure to accurately correct parallax systematics to within $\sim 1\,\mu$as \citep{flynn2022clusters,apellaniz2022clustersclouds} in the optimal magnitude and color range established above. As a result, a significant nonzero residual parallax offset is expected for (bright) Cepheids, while no residual parallax offset is expected for cluster members after the L21 corrections are applied.

The final cluster parallaxes were computed as the weighted mean of the retained cluster members. The total parallax uncertainty sums  the  statistical uncertainty determined as the error on the weighted mean in quadrature and the systematic contribution due to angular covariance determined by \citet{apellaniz2021validation}. Because our initial search radius around Cepheids is $1$\,deg, the full diameters of all clusters is significantly smaller than $2$\,deg.  This allowed us to consider the  estimation of angular covariance based on the LMC alone as given by $V_{\varpi, \mathrm{LMC}}$ in their Eq. 10 \citep[cf. also Sect. 2.2 in][]{2022arXiv220606212R}, neglecting
wide-angle contributions estimated using quasars. This is analogous to the approach taken by \citet{zinn2021validation} in conjunction with the angular covariance estimates based on the \textit{Kepler} field. In practice, this reduces the error floor for average cluster parallaxes from $\lesssim 10$ to $\lesssim 7\,\mu$as. Because the mean separation of our Cepheid clusters is very large, covariance among clusters is negligible.

\subsubsection{Maximum angular separations }\label{sec:Separation}

We calculated the projected distance of the Cepheid from  cluster center assuming that both objects are at the distance of the cluster. Candidate associations with separations greater than $25$\,pc were discarded in favor of sample purity and to ensure that cluster average parallaxes can be used as accurate proxies for Cepheid parallaxes. Hypothetical Cepheids residing in extended tidal tails \citep{Jerabkova2021} would thus be excluded from our analysis. We refer to Cepheids as coronal cluster members if their projected separation from cluster center exceeds $8$\,pc but does not exceed $25$\,pc.

\subsubsection{Proper motions}\label{sec:ClPM}
We computed bulk cluster proper motions as the mean of all clusters members as well as proper motion dispersions using cluster members in the color and magnitude range used for parallaxes. We used proper motions to reject cluster candidates as spurious asterisms if the resulting velocity dispersion exceeded reasonable values for gravitationally bound systems following  \citet{cantat2020clusters} and \citet{hunt2021improving}. Specifically, up to  $\varpi=0.67$\,mas, we rejected associations whose projected velocity dispersion exceeds $5 \times \sqrt{2}\, \mathrm{[mas/yr]}$ ($5$\,\kms), whereas a maximum difference of $1\,\text{mas yr}^{-1}$ was allowed for clusters with a smaller parallax to reflect the increased uncertainties, in particular, of the fainter main-sequence cluster members. Thus, we required\footnote{\label{foot:pm}We recall that $\mu \mathrm{[mas/yr]} \approx v \mathrm{[km/s]} \cdot \varpi \mathrm{[mas]} /4.74$, so that $1\,\mathrm{[mas/yr]} \approx 5 \mathrm{[km/s]} \times \sqrt{2}$ at $\varpi=0.67$\,mas (1.5 kpc).}
\begin{align}\label{eq:1}
    \sqrt{\sigma_{\mu_{\alpha}^{*}}^{2} + \sigma_{\mu_{\delta}}^{2}}\leq \left\{\begin{matrix}
1 \quad \text{mas yr}^{-1} \quad \text{if}\quad \varpi \leq 0.67 \quad \text{mas}\\ 
5\sqrt{2} \frac{\varpi}{4.74} \quad \text{mas yr}^{-1} \quad \text{if} \quad \varpi > 0.67 \quad \text{mas}.
\end{matrix}\right.
\end{align}
In practice, however, all retained clusters exhibit a significantly lower velocity dispersion, with a mean value of $ 2.8$\,\kms (cf. Fig.\,\ref{app:PMdispersion}).
Inspection revealed that the proper motion dispersion estimated using only spatially densely concentrated cluster members returned by our clustering analysis underestimated the intrinsic velocity dispersion of true cluster members observed at large angular separations, which  require a statistically greater velocity dispersion to reach their large separations from cluster centers. To avoid unrealistically low membership probabilities for coronal cluster Cepheids (cf. Sec.\,\ref{sec:likelihood}), we therefore adopted twice the standard deviation determined based on the member stars recovered by HDBSCAN as the more conservative estimate of true cluster proper motion dispersion when assessing Cepheid membership in clusters.

\subsubsection{Radial velocity}\label{sec:RV}

Cluster radial velocities (RV) are computed using \gaia\ DR3  mean radial velocities \citep[parameter \texttt{radial\_velocity} from table \texttt{gaia\_source}]{gaiadr3.radvel}. For each cluster with available DR3 RVs, Table\,\ref{tab:RV} lists the number of (non-Cepheid) cluster member stars, their median RV, standard error on the cluster median RV, and the Cepheid paramaters. 

We did not consider cluster RVs based on few stars ($\lesssim 3$) sufficiently reliable for further analysis. Thus, we did not consider RV as a membership constraint for the candidate host clusters of WX~Pup, CV~Mon, IQ~Nor, and SX~Vel.

\subsection{Cepheid membership determination} \label{sec:likelihood}

We computed cluster membership probabilities for Cepheids whose proper motions and parallaxes separately agreed to within approximately $3\sigma$ of their potential host cluster parameters. This subsection presents our method, and the resulting probabilities are presented in Sec.\,\ref{sec:CC}. A tolerance of up to $0.5\sigma$ was permitted in this initial screening. In this context, $\sigma$ refers to combined (square-summed) dispersions or uncertainties, depending on the parameter, of clusters and Cepheids as follows. For proper motions, the cluster dispersion as described in Sec.\,\ref{sec:ClPM} was combined with the Cepheid uncertainties reported by \gaia. For parallaxes, $\sigma$ contains the squared sum of uncertainties of the weighted cluster average (no significant internal dispersion expected), the individual Cepheid parallax uncertainty, and an additional $15\,\mu$as uncertainty to reflect the magnitude dependence of the residual parallax offset after applying the L21 corrections.

We computed Cepheid membership probabilities using the likelihood formalism developed in \citet{anderson2013cepheids} and the membership constraints $\varpi$, $\mu_{\alpha}^{*}$, $\mu_\delta$, and RVs. Strictly speaking, this approach performs a hypothesis test under the null hypothesis of the Cepheid cluster membership and can only reject this null hypothesis, not prove it. As in \citet{anderson2013cepheids}, we computed the Bayesian likelihood $P(B|A) = 1 - p(c)$, where $p(c)$ is the $p-$value of the $\chi^2_{N_{\mathrm{dof}}}$ distributed quantity
\begin{equation}
    c = \mathbf{x^T \Sigma^{-1} x}
,\end{equation}
where the vector $x$ contains the differences between Cepheid and cluster parameters, that is,
\begin{equation}
  x = (\varpi_{\mathrm{Cl}}-\varpi_{\mathrm{Cep}},v_{r,\mathrm{Cl}} - v_{\gamma,\mathrm{Cep}}, ...  ) \ ,
\end{equation}
and $\Sigma$ is the diagonal covariance matrix containing the squared values of $\sigma$ for the various membership constraints, as explained above. Our threshold for rejecting the membership hypothesis was $P(B|A) < 0.0027$, which corresponds to a $3\sigma$ rejection criterion. Stars with a higher probability are considered bona fide cluster Cepheids provided the host cluster detection is sufficiently robust.

Radial velocities were included in this calculation if cluster average RVs ($v_{r,\mathrm{Cl}}$) could be estimated using at least three  member stars and if Cepheid systemic radial velocities, $v_\gamma$, could be determined using a Fourier series fit to time-series data from either the velocities of Cepheids project (cf. Anderson et al. in prep., VELOCE~I) or the literature \citep[e.g.,][]{anderson2016vetting}. In addition to cluster average values, Table\,\ref{tab:RV} lists RV data for Cepheids, including $v_\gamma$, its uncertainty, references to data used, and the difference between cluster median and Cepheid $v_\gamma$, the total uncertainty (summed in quadrature), and the difference between cluster and Cepheid in units of the total uncertainty. The only Cepheid for which RV information significantly contradicts membership is XZ~Car, which is part of our Silver sample (cf. Sect.\,\ref{sec:XZCar}). All other stars are found to agree to within $1.35\sigma$ with their host cluster median velocities. Further information about Cepheid RVs and \gaia\ DR3 radial velocities of Cepheids will be provided as part of the VELOCE project (Anderson et al. in prep.).

\begin{table*}
\centering
\caption{Radial velocity information for clusters and Cepheids\label{tab:RV}}
\begin{tabular}{lrcrlccrccr}\toprule
Cluster & $N_{\mathrm{RV}}$ & median RV & $\sigma_{\mathrm{RV,Cl}}$ & Cepheid & $v_\gamma$ & $\sigma_{v_\gamma}$ & Refs & $\Delta_{\mathrm{Cl-Cep}}$ & $\sigma_{\mathrm{tot}}$ & $N_\sigma$ \\ 
 & & (\kms) & (\kms) &  & (\kms) & (\kms) & & (\kms) & (\kms) &  \\ \midrule
UBC 129 & 9 & -14.84 & 6.18 & X Vul & -14.54 & 0.14 & a;b;c & -0.30 & 6.18 & 0.05  \\
NGC 103 & 7 & -77.88 & 9.89 & NO Cas & -79.13 & 0.17 & V & 1.25 & 9.89 & 0.13  \\
UBC 130 & 3 & -0.20 & 2.08 & SV Vul & -0.68 & 2.80 & V & 0.49 & 3.49 & 0.14  \\
NGC 6067 & 93 & -39.04 & 3.10 & QZ Nor & -38.60 & 0.13 & V & -0.44 & 3.10 & 0.14  \\
UBC 290 & 19 & -23.50 & 5.17 & X Cru & -24.26 & 0.15 & d & 0.76 & 5.17 & 0.15  \\
NGC 6067 & 93 & -39.04 & 3.10 & V0340 Nor & -39.63 & 0.09 & V & 0.59 & 3.10 & 0.19  \\
Berkeley 58 & 3 & -71.36 & 28.59 & CG Cas & -77.52 & 0.56 & e;c;V & 6.16 & 28.59 & 0.22  \\
UBC 533 & 9 & -17.96 & 7.95 & GH Lup & -16.11 & 0.25 & f & -1.85 & 7.95 & 0.23  \\
UBC 375 & 6 & -13.13 & 9.86 & V0438 Cyg & -10.06 & 2.05 & e & -3.07 & 10.07 & 0.30  \\
FSR 0951 & 16 & 44.02 & 3.05 & RS Ori & 42.94 & 0.04 & V & 1.08 & 3.05 & 0.35  \\
Cl X Pup & 4 & 74.70 & 10.07 & X Pup & 71.05 & 0.38 & V & 3.65 & 10.08 & 0.36  \\
NGC 6649 & 21 & -1.44 & 24.70 & V0367 Sct & -10.60 & 1.26 & e;g;b;c & 9.16 & 24.73 & 0.37  \\
Lynga 6 & 14 & -46.98 & 24.41 & TW Nor & -56.60 & 0.29 & g;b & 9.62 & 24.41 & 0.39  \\
NGC 129 & 5 & -52.12 & 6.51 & DL Cas & -46.53 & 0.32 & V & -5.59 & 6.52 & 0.86  \\
NGC 6664 & 12 & 9.33 & 8.54 & Y Sct & 14.03 & 0.15 & V & -4.70 & 8.54 & 0.55  \\
IC 4725 & 184 & 2.03 & 1.71 & U Sgr & 3.16 & 0.09 & V & -1.13 & 1.71 & 0.66  \\
CWNU 175 & 4 & 6.19 & 11.53 & VW Cru & -2.82 & 5.30 & h & 9.01 & 12.69 & 0.71  \\
NGC 7790 & 6 & -68.61 & 11.94 & CF Cas & -77.76 & 0.15 & e;h;b;c & 9.15 & 11.94 & 0.77  \\
UBC 106 & 28 & 41.88 & 2.84 & CM Sct & 39.63 & 0.24 & e;g;h & 2.25 & 2.85 & 0.79  \\
NGC 6087 & 69 & 2.91 & 2.67 & S Nor & 6.09 & 2.12 & V & -3.18 & 3.41 & 0.93  \\
Cl ST Tau & 15 & 9.01 & 6.79 & ST Tau & 1.73 & 0.26 & a;b;c;i;V & 7.28 & 6.80 & 1.07  \\
Cl V0378 Cen & 7 & -25.57 & 8.50 & V0378 Cen & -16.24 & 0.04 & V & -9.33 & 8.50 & 1.10  \\
Ruprecht 79 & 11 & 34.55 & 6.28 & CS Vel & 27.10 & 0.19 & g;b & 7.45 & 6.29 & 1.19  \\
NGC 5662 & 89 & -26.60 & 3.59 & V Cen & -21.74 & 0.08 & V & -4.86 & 3.59 & 1.35  \\
UBC 231 & 3 & 75.64 & 7.54 & WX Pup & 52.98 & 0.03 & V & 22.66 & 7.54 & (3.01)  \\
Ruprecht 93      & 13 & -27.77 & 2.73 & XZ Car & 5.72 & 0.42 & V & -33.49 & 2.76 & 12.12  \\
vdBergh 1 & 1 & 9.35 & 0.00 & CV Mon & 19.44 & 0.11 & a;g;c;j & -10.09 & 0.11 & (91.71)  \\
Cl IQ Nor & 1 & 6.07 & 0.00 & IQ Nor & -24.51 & 0.13 & V & 30.58 & 0.13 & (244.66)  \\
Cl SX Vel & 1 & -22.52 & 0.00 & SX Vel & 29.79 & 0.05 & V & -52.31 & 0.05 & (1046.18)  \\ \midrule
\end{tabular}
\tablefoot{RV differences between clusters and Cepheids are considered significant only if a sufficient number (here: $\gtrsim 3$) of cluster stars was available to determine an accurate median for the cluster. The last column shows apparently highly discrepant values in parentheses if they are based on an insufficient number of stars. References listed in Column `Refs' a: \citet{1988ApJS...66...43B}, b: \citet{1994A&AS..108...25B}, c: \citet{1992SvAL...18..316G}, d: \citet{2002ApJS..140..465B}, e: \citet{1991ApJS...76..803M}, f: \citet{1985SAAOC...9....5C}, g: \citet{1992AJ....103..529M}, h: \citet{1994A&AS..105..165P}, i: \citet{1999A&AS..140...79I}, j: \citet{2004A&A...415..531S}, V: VELOCE (Anderson et al. in prep.)}
\end{table*}

In contrast to \citet{anderson2013cepheids}, we did not explicitly use the angular separation as an external multiplicative prior because individual cluster members were already separated from the background by our clustering analysis. However, our use of a maximum allowed projected separation of $25$\,pc could be seen as a flat prior with $P(A) = 1$ for absolute projected separations smaller than this cutoff value. Ages and chemical compositions were not considered in the calculation of the likelihood.

\section{Cluster Cepheids}\label{sec:CC}

\label{sec:CCsamples}
We grouped our sample of cluster Cepheids into Gold, Silver, and Bronze samples according to the following criteria. The Gold sample contains cluster Cepheids whose host cluster detections were robust and whose membership likelihoods exceeded the threshold for rejecting the membership hypothesis (cf. Sec.\,\ref{sec:likelihood}). This sample is best suited for LL calibration. The Silver sample contains cases where the host cluster detection is solid, whereas the likelihood computation quantitatively rejects cluster membership due to a difference slightly larger than $3\sigma$  in individual constraints. This sample is of particular interest for the further study to refine possible membership, for instance, taking uncertainties related to stellar multiplicity into account. The Bronze sample is composed of two cases for which the host cluster detection is not as clean as in the Gold sample.

Tables \ref{tab:Gold}-\ref{tab:Copper} list the Cepheids and their host clusters for the Gold, Silver, and Bronze samples, along with their main astrometric information.  Representative examples of each  set are shown in Figs.\,\ref{fig:examples1} and \ref{fig:examples2}. We applied an additional uncertainty of $15\,\mu$as when we computed the significance of the disagreement in parallax (cf. Sec.\,\ref{sec:likelihood}). Table\,\ref{tab:GaiaSample2} provides a list of the Gaia EDR3 source~ids for all the cluster members and their  L21 corrected parallaxes.

\subsection{Gold sample\label{sec:Gold}}

\begin{sidewaystable*}{}
  \centering
   \ra{1.2}

\begin{tabular}{lrrrrrrclrrrrcr}\toprule
 \multicolumn{7}{c}{Cluster parameters} & \phantom{}& \multicolumn{6}{c}{Cepheid parameters}\\
\cmidrule(lr){0-6} \cmidrule(lr){8-14} 
Cluster & $\alpha (^{\circ})$  & $\delta (^{\circ})$ & N &  $\varpi$ ($\mu$as) & $\mu^{*}_{\alpha}$ (mas/yr)  & $\mu_{\delta}$ (mas/yr)  && Cepheid& $\varpi$ ($\mu$as)  & $\mu^{*}_{\alpha}$ (mas/yr)  & $\mu_{\delta}$ (mas/yr) & Sep (pc) & M.P\\ \midrule
        Czernik 41 & $ 297.746$ & $ 25.279$ & 112 & $ 407 \pm 8$  & $-2.949 \pm 0.102$  &  $-6.180 \pm 0.104$   && ATO J297$\dagger^{(*)}$ &  $340 \pm 18 $  & $-3.025 \pm 0.012$& $-6.294 \pm 0.017$ &  0.82  & 0.05 \\
        NGC 7790 & $359.619  $ & $61.206$ & 149 & 322 $\pm$ 7 & -3.229 $\pm$ 0.115 & -1.729 $\pm$ 0.085 && CE Cas A & 332 $\pm$ 15 & -3.298 $\pm$ 0.015 & -1.873 $\pm$ 0.017 & 2.3 & 0.80  \\  
        NGC 7790 & $359.619  $ & $61.206$ & 149 & 322 $\pm$ 7 & -3.229 $\pm$ 0.115 & -1.729 $\pm$ 0.085 && CE Cas B & 333 $\pm$ 15 & -3.301 $\pm$ 0.014 & -1.809 $\pm$ 0.016 & 2.3 &  0.90 \\ 
        NGC 7790 & $359.619  $ & $61.206 $ & 149 & 322 $\pm$ 7 & -3.229$\pm$ 0.115 & -1.729 $\pm$ 0.085 && CF Cas &  316 $\pm$ 12 & -3.240 $\pm$ 0.012 & -1.766 $\pm$ 0.012 & 1.5&  0.95 \\  
        Berkeley 58 & $0.076$ & $60.947$& 183 & 336 $\pm$ 7  & -3.430 $\pm$ 0.237 & -1.791 $\pm$ 0.130 && CG Cas & 296 $\pm$ 14 & -3.241 $\pm$ 0.013 & -1.673 $\pm$ 0.015 & 4.8& 0.43 \\  
        UBC 106 & $280.469      $ & $-5.411 $& 495& 443 $\pm$ 7  & -1.048 $\pm$ 0.108 & -1.365 $\pm$ 0.134 && CM Sct & 444 $\pm$ 16 & -1.064 $\pm$ 0.015 & -1.414 $\pm$ 0.014 & 6.7 & 0.96\\  
        Ruprecht 79 & $145.253$ & $-53.850$& 152 & 281 $\pm$ 7  & -4.615 $\pm$ 0.183 & 3.086 $\pm$ 0.200 && CS Vel & 272 $\pm$ 12 & -4.567 $\pm$ 0.014 & 3.131 $\pm$ 0.014 & 2.7& 0.81 \\
        vdBergh 1 & $ 99.283 $ & $3.074 $&60 & 585 $\pm$ 10 & 0.411 $\pm$ 0.204 & -0.704 $\pm$ 0.185 && CV Mon & 601 $\pm$ 15 & 0.349 $\pm$ 0.016 & -0.666 $\pm$ 0.014 & 0.5&   0.92    \\  
        NGC 129 & $7.606 $ & $60.200 $ & 297& 556 $\pm$ 7  & -2.594 $\pm$ 0.120 & -1.169 $\pm$ 0.104 && DL Cas & 580 $\pm$ 27 & -2.706 $\pm$ 0.025 & -1.189 $\pm$ 0.027 & 1.9& 0.82\\ 
        NGC 6664 & $279.115     $ & $-8.190$& 361 & 504 $\pm$ 7  & -0.065 $\pm$ 0.176 & -2.514 $\pm$ 0.351 && EV Sct$^{(*)}$ & 526 $\pm$ 18 & -0.209 $\pm$ 0.018 & -2.546 $\pm$ 0.015 & 1.9&  0.79 \\ 
        UBC 533 & $230.311$ & $-53.176$ & $68$ & $878 \pm 9$ & $-1.766 \pm 0.192$ & $-1.381 \pm 0.207$ && GH Lup  &  $864 \pm 21$ & $-1.337 \pm 0.021$ & $-2.202 \pm 0.020$ & 12.7 & 0.25 \\
        Cl IQ Nor & $228.444$ & $-54.590$& 48 & 544 $\pm$ 9 & -0.999 $\pm$ 0.433 & -1.790 $\pm$ 0.173 && IQ Nor & 535 $\pm$ 18 & -0.897 $\pm$ 0.015 & -1.821 $\pm$ 0.020 & 7.4&     0.99 \\  
        NGC 103& $6.391 $ & $61.315 $ &  243& 317 $\pm$ 7  & -2.695 $\pm$ 0.383 & -1.031 $\pm$ 0.149 && NO Cas$^{(*)}$ & 298 $\pm$ 13 & -2.828 $\pm$ 0.012 & -1.208 $\pm$ 0.012 & 11.3& 0.87\\ 
        NGC 6067& $243.304      $ & $-54.229$& 1085 & 513 $\pm$ 7 & -1.948 $\pm$ 0.149 & -2.595 $\pm$ 0.186 && QZ Nor$^{(*)}$ & 484 $\pm$ 20 & -1.896 $\pm$ 0.023 & -3.848 $\pm$ 0.019 & 11& 0.01 \\  
        FSR 0951& $95.572$ & $14.635 8$& 176 & 610 $\pm$ 7 & 0.214 $\pm$ 0.102 & 0.032 $\pm$ 0.142 && RS Ori & 589 $\pm$ 30 & 0.196 $\pm$ 0.036 & 0.005 $\pm$ 0.028 & 1.5 & 0.97 \\  
        NGC 6087& $244.742       $ & $-57.912$ & 196 & $1073 \pm 7 $ & -1.603 $\pm$ 0.272 & -2.424 $\pm$ 0.247 && S Nor & 1099 $\pm$ 22 & -1.608 $\pm$ 0.025 & -2.136 $\pm$ 0.020 & 0.3 & 0.72 \\  
        Cl ST Tau& $ 85.740 $ & $13.719$ & 79 & 953 $\pm$ 8 & 0.691 $\pm$ 0.248 & -3.636 $\pm$ 0.288 && ST Tau & 916 $\pm$ 34 & 0.188 $\pm$ 0.035 & -2.318 $\pm$ 0.022 & 10.0 & 0.08 \\
        UBC 130& $298.043 $ & $27.443 $& 42 & 425 $\pm$ 9 & -2.103 $\pm$ 0.071 & -5.872 $\pm$ 0.102 && SV Vul & 402 $\pm$ 21 & -2.158 $\pm$ 0.016 & -5.962 $\pm$ 0.021 & 6.5 &  0.90 \\  
        Cl SX Vel& $131.268 $ & $-46.095$& 77 & 497 $\pm$ 7 & -5.057 $\pm$ 0.160 & 4.976 $\pm$ 0.120 && SX Vel & 501 $\pm$ 19 & -4.345 $\pm$ 0.019 & 4.921 $\pm$ 0.022 & 9.6 & 0.17\\ 
        Lyngå 6& $241.209 $ & $-51.952$& 173 & 421 $\pm$ 8 & -1.915 $\pm$ 0.187 & -2.762 $\pm$ 0.183 && TW Nor & 360 $\pm$ 20 & -1.891 $\pm$ 0.021 & -2.806 $\pm$ 0.017 & 0.6&  0.25\\ 
        IC 4725& $277.958       $ & $-19.126 $& 492 & 1554 $\pm$ 6 & -1.685 $\pm$ 0.239 & -6.159 $\pm$ 0.319 && U Sgr & 1605 $\pm$ 22 & -1.795 $\pm$ 0.025 & -6.127 $\pm$ 0.017 & 0.1& 0.42\\  
        NGC 5662& $218.731      $ & $-56.665$& 241 & 1337 $\pm$ 6  & -6.461 $\pm$ 0.157 & -7.189 $\pm$ 0.180 && V Cen & 1409 $\pm$ 22 & -6.697 $\pm$ 0.016 & -7.068 $\pm$ 0.018 & 17.2 & 0.05\\
        FSR 0384& $342.921      $ & $56.106      $& 63 & 520 $\pm$ 8 & -3.446 $\pm$ 0.198 & -1.752 $\pm$ 0.096 && V Lac & 496 $\pm$ 16 & -3.237 $\pm$ 0.016 & -1.439 $\pm$ 0.017 & 16.9 & 0.27\\ 
        NGC 6067& $243.304      $ & $-54.229$& 1085 & 513 $\pm$ 7 & -1.948 $\pm$ 0.149 & -2.595 $\pm$ 0.186 && V0340 Nor & 491 $\pm$ 25 & -2.066 $\pm$ 0.027 & -2.634 $\pm$ 0.021 & 0.5 & 0.94\\  
        NGC 6649& $278.359$ & $-10.399  $&  425 & 514 $\pm$ 7 & 0.025 $\pm$ 0.132 & -0.121 $\pm$ 0.156 && V0367 Sct & 473 $\pm$ 20 & 0.082 $\pm$ 0.021 & -0.273 $\pm$ 0.019 & 1.7& 0.60\\  
        Cl V0378 Cen & $199.642 $ & $-62.593$ &107 & 518   $\pm$ 8 & -4.325 $\pm$ 0.319 & -1.744 $\pm$ 0.194 && V0378 Cen$^{(*)}$ & 524 $\pm$ 19 & -5.656 $\pm$ 0.014 & -2.282 $\pm$ 0.019 & 7.7 & 0.11 \\
        NGC 129& $7.608 $ & $60.199 $ & 311& 557 $\pm$ 7  & -2.591 $\pm$ 0.123 & -1.169 $\pm$ 0.113 && V0379 Cas$^{(*)}$ & 524 $\pm$ 14 & -2.696 $\pm$ 0.012 & -1.313 $\pm$ 0.015 & 25.0&  0.42\\  
        UBC 375& $304.617 $ & $40.062 $& 160 & 562 $\pm$ 7 & -3.073 $\pm$ 0.461 & -5.219 $\pm$ 0.284 && V0438 Cyg & 530 $\pm$ 16 & -3.324 $\pm$ 0.017 & -4.559 $\pm$ 0.019 & 2.9& 0.48 \\  
        CWNU 175& $188.395      $ & $-63.506$& 37 & 732 $\pm$ 9 & -3.987 $\pm$ 0.277 & -1.187 $\pm$ 0.165 && VW Cru & 738 $\pm$ 16 & -3.903 $\pm$ 0.015 & -1.134 $\pm$ 0.015 & 1.1& 0.96 \\
        UBC 231 & $115.571 $ & $-25.265 $& 68 &  $345 \pm 8 $  &  $-2.177 \pm 0.240 $  &  $2.297 \pm 0.340$   && WX Pup &  387 $\pm$ 15  &  -2.163 $\pm$ 0.010 & 2.559 $\pm$  0.014 & 22.2 & 0.01  \\  
        UBC 290& $191.807       $ & $-59.373$& 253 & 639 $\pm$ 6 & -5.918 $\pm$ 0.279 & -0.273 $\pm$ 0.198 && X Cru & 654 $\pm$ 19 & -5.926 $\pm$ 0.015 & -0.173 $\pm$ 0.017 & 7.6&  0.98\\ 
        FSR 0384& $342.921      $ & $56.106      $& 63 & 520 $\pm$ 8 & -3.446 $\pm$ 0.198 & -1.752 $\pm$ 0.096 && X Lac$^{(*)}$ & 520 $\pm$ 18 & -3.296 $\pm$ 0.017 & -1.442 $\pm$ 0.017 & 17.2 &  0.43\\
        UBC 129& $299.131$  & $26.464$& 131 & 880 $\pm$ 7 & -1.031 $\pm$ 0.138 & -4.363 $\pm$ 0.189 && X Vul & 864 $\pm$ 22 & -1.352 $\pm$ 0.016 & -4.247 $\pm$ 0.020 & 4.9& 0.77 \\  
        NGC 6664& $279.115      $ & $-8.190$& 361 & 504 $\pm$ 7  & -0.065 $\pm$ 0.176 & -2.514 $\pm$ 0.351  && Y Sct & 558 $\pm$ 20 & -0.737 $\pm$ 0.025 & -2.878 $\pm$ 0.019 & 16.23 & 0.08 \\

      \bottomrule
\end{tabular}
\caption{Gold sample of cluster Cepheids.
\textit{Left:} Host cluster parameters. \textit{Right:} Cepheid parameters. The average cluster parallaxes were estimated using stars in the range $12.5<G<17$ as explained in Sect.~\ref{sec:CC}. The uncertainty includes the contribution from angular covariance. $^{(*)}$ denotes first overtone pulsators. The second last column states the projected separation of the Cepheid from cluster center in pc. The last column states the membership probability if HDBSCAN considers the Cepheid a member and "-" if not. ATO J297$\dagger$ abbreviates the full identifier of ATO~J297.7863+25.3136.}
\label{tab:Gold}
\end{sidewaystable*}

The Gold sample consists of 34 Cepheids residing in 28 distinct Galactic open clusters. Out of the 34 Cepheids, 27 Cepheids pulsate in the fundamental mode, and 7 Cepheids pulsate in the first overtone. We identify ST~Tau, V0378~Cen, and GH~Lup as bona fide cluster Cepheids for the first time.

We cross-matched all 28 Gold sample host clusters with cluster catalogs from the literature \citep{anderson2013cepheids,usenko2019spectroscopic,cantat2018gaia,2020A&A...640A...1C,he2022new,hunt2021improving,medina2021revisited}. We found cluster parameters in agreement to within $1 \sigma$  of the previously reported parameters in the literature for 24 of them.  However, we found disagreements greater than $2\sigma$ among at least one of the astrometric parameters for the host clusters of SX~Vel, IQ~Nor, and VW~Cru.

Last but not least, we  identified four entirely new clusters that host one Cepheid each. We denoted them by the prefix \textit{Cl} followed by the Cepheid name. Additional information for a subset of Gold sample cluster Cepheids is provided below.

\paragraph{SX Vel}\label{sec:sxvel} is found to be a member of a newly detected host cluster (Cl SX~Vel, $d=2012 \pm 29$ pc) at a projected separation of $9.6\,$pc. The presence of multiple clusters in close proximity somewhat complicates this membership analysis. \citet{anderson2013cepheids} investigated multiple possible host clusters, including  Bochum 7, NGC 2660, FSR 1441, SAI 94, and Ruprecht 70, and we here add NGC\,2659. Membership in Bochum 7 \citep[$5754$\,pc; cf.][]{kharchenko2005astrophysical} and SAI 94  \citep[$3515\pm 60$ pc; cf.][]{elsanhoury2019photometric} is readily excluded based on distance, while proper motion differences exclude membership in NGC~2660, and FSR~1441 \citep{cantat2018gaia}. However, NGC\,2659 and Ruprecht\,70 require some discussion because the computed likelihoods for cluster membership are consistent with the hypothesis of membership for both and because the likelihood obtained for NGC\,2659 is even higher ($0.65$) than for Cl SX~Vel ($0.17$). However, closer inspection revealed that the higher likelihood for NGC\,2659 is driven by weaker proper motion constraints (twice larger dispersion). The parallaxes of both clusters agree to within $1.1\sigma$ ($497 \pm 7\,\mu$as vs $508\pm 7\,mu$as). Additionally, the observed separation of $43$\,pc is inconsistent with our maximum allowed separation of $25\,$pc. Similarly, for Ruprecht\,70, the separation of $34$\,pc rejects this association, although the likelihood alone ($0.004$) would not reject membership according to our criteria.

\paragraph{IQ Nor}\label{sec:iqnor} is associated with a cluster at a distance of   $1839 \pm 32$ pc. Previously, \cite{anderson2013cepheids} investigated the membership of IQ Nor in the following clusters: NGC 5582, NGC 5925, or Loden 2115. Membership in NGC~5822 is excluded based on parallax \citep[$12\sigma$ difference][]{cantat2018gaia}, whereas proper motions exclude membership in NGC~5925 \citep{cantat2018gaia} and Loden~2115 \citep{kharchenko2013global}.

\paragraph{VW Cru}\label{sec:vwcru} resides in a cluster reported independently as CWNU~175 while this article was in preparation \citep{he2022new}. 
Although \citet{anderson2013cepheids} previously investigated possible membership in Loden~624 \citep{kharchenko2013global}, we note that CWNU~175 is a different physical object separated by $1.9$\,deg from Loden~624.

\paragraph{WX Pup\label{sec:WXPup}} is a coronal member of the cluster UBC~231 \citep[see also][]{zhou2021galactic} and a good example of how the \gaia\  systematics can limit the ability to detect host clusters because the cluster and Cepheid parallaxes differed by $3.6\sigma$ prior to applying L21 parallax corrections. After applying L21 corrections, this differences reduces to $1.8\sigma$. While the membership likelihood of WX Pupis a relatively low $1\%$ and the projected separation of 22.2\,pc is close to our cutoff, its membership in UBC~231 is not rejected according to the criteria we specified. We searched for other cases where the chronological order of the L21 corrections would affect the conclusion concerning membership, but found none.

\paragraph{ATO J297.7863+25.3136} was discovered recently \citep{heinze2018first} and identified as a member of Cluster~41 by \citet{medina2021revisited}. We here confirm this association at a distance of $2456 \pm 49$ pc. However, this cluster is located in a highly reddened region of the sky, limiting its usefulness for LL calibration (cf. Fig.\,\ref{fig:ATLAS}).

\paragraph{SV Vul} is especially valuable for LL calibration due to its long period because the majority of Cepheids in distant supernova-host galaxies have periods $\log{P} > 1.2$ \citep[e.g.,][]{riess2018new}. We find a very high likelihood of $90\%$ for this cluster Cepheid combination at a distance of $ 2354  \pm 49$ pc, and we note the small $\sim 6.5$\,pc separation from cluster center. Thus, our analysis confirms previous statements of the SV~Vul cluster membership reported by \cite{negueruela2020cluster} and \cite{medina2021revisited}. Moreover, inspection of several membership constraints as well as the residuals from  our LL calibration does not corroborate the possibility that the parallaxes of SV~Vul are unreliable, reported by \citet{owens2022current} (cf. Fig.\,\ref{fig:ABL_optical} and Sect.\,\ref{sec:combined}). We therefore find no reason to discard this valuable star from LL calibration.

\subsection{Silver  sample\label{sec:silver}}

The Silver sample contains three Cepheids with likelihoods that are formally inconsistent with membership in well-defined clusters according to our criteria. However, disagreements among the individual membership constraints are sufficiently small to warrant additional discussion and inspection.

\paragraph{AP Vel}\label{sec:APVel} was previously reported as a member of the cluster Ruprecht 65 \citep{chen2015search} located at a distance of $ 2085 \pm 32$ pc. The low membership probability is dominated by the $3.3\sigma$ parallax difference.    We do note, however, that the proper motion parameters of AP\ Vel ($\mu_\alpha^*$, $\mu_\delta$) are within $2.3$ and $1.7\sigma$ of the cluster averages, and that the Cepheid is located rather close to (0.21 deg) from cluster center. 

\paragraph{X Pup}\label{sec:XPup} is located at a rather large separation of $\sim 24.3$ pc from the center of its  possible newly identified host cluster. The low likelihood is driven by proper motion differences between Cepheid and cluster, which are significant at the level of $\sim 3.1$ and $3.3\sigma$ for $\mu_\alpha^*$ and $\mu_\delta$, respectively. However, we note that the total velocity dispersion of Cl X~Pup is merely $3.3\,$\kms, which may indicate that an underestimated proper motion dispersion was used to calculate the membership. Additionally, the comparatively large separation from the cluster (cf. Sec.\,\ref{sec:ClPM}) as well as orbital motion tentatively reported by \cite{anderson2016vetting} may contribute to deviations in proper motion. We note the good agreement in parallax ($1.4\sigma$) and radial velocity, where the Cepheid barycentric velocity is $ 71.02 \pm 0.16 $ km/s  (Anderson et al. in prep), which is fully consistent with  the median cluster radial velocity based on four stars reported in \gaia\ DR3 ($ 74 \pm 10$ km/s) (cf. Table\,\ref{tab:RV}). We therefore consider the cluster membership of X Pup to be potentially underestimated due to an underestimated cluster proper motion dispersion. Further study is required to ascertain its membership before X~Pup is included in the Gold sample.

\paragraph{XZ Car\label{sec:XZCar}} is situated at a projected separation of 15\,pc from its potential newly identified host cluster Ruprecht\,93. Although the parallax of XZ~Car fully agrees with that of the  cluster, we find a low membership probability due to differences in the kinematic membership constraints, notably radial velocities, which  differ by $33\,$\kms\ between the pulsation-averaged Cepheid RV and the median RV of the 13 cluster members with DR3 radial velocities (cf. Table\,\ref{tab:RV}). Although XZ Car is a long-term spectroscopic binary and exhibits a trend of its pulsation-averaged velocity $v_\gamma$ that exceeds $5$\kms\ over a baseline of $\sim 40$\,yr \citep[Shetye et al. in prep.]{anderson2016vetting}, we caution that orbital motion is unlikely to explain the large RV difference. Additionally, $\mu_\alpha^*$ and $\mu_\delta$ differ by $2.9\sigma$ and $2.6\sigma$. We note that evidence of orbital motion has also been found using \gaia\ proper motion anomalies \citep{kervella2019multiplicity}, however, indicating that proper motion may also provide incorrect membership indications for XZ~Car. It would be intriguing (but beyond scope for this article) to investigate the nature of the orbit and the companion required to explain these differences. However, XZ~Car does not appear to be gravitationally bound to Ruprecht\,93. Further membership analysis using the full \gaia\ temporal baseline might clarify this high-interest association.

\begin{sidewaystable*}{}
     \ra{1.3}
   \centering
\begin{tabular}{lrrrrrrclrrrrr}\toprule
 \multicolumn{4}{c}{Cluster parameters} & \phantom{   }& \multicolumn{5}{c}{Cepheid parameters}\\
\cmidrule(lr){0-6} \cmidrule(lr){9-14} 
Cluster& $\alpha (^{\circ})$  & $\delta (^{\circ})$ & N  & $\varpi$ ($\mu$as) & $\mu^{*}_{\alpha}$ (mas/yr)  & $\mu_{\delta}$ (mas/yr)  && Cepheid& $\varpi$ ($\mu$as) & $\mu^{*}_{\alpha}$ (mas/yr)  & $\mu_{\delta}$ (mas/yr) &  Sep  (pc) &  M.P \\ \midrule
      Ruprecht 65 & $129.818 $ & $ -44.062$ &  133 & $ 478 \pm 7$  &  $-4.795 \pm 0.270$  &  $4.424 \pm 0.360$  && AP Vel$^{(*)}$ & 545 $\pm$ 12 & -6.030      $\pm$ 0.013 &  5.647 $\pm$  0.013 & 8.5  & 0 \\ 
        Ruprecht 93      & $166.029  $ & $-61.371$&  206 & 482 $\pm$ 7 & -6.437 $\pm$ 0.143 & 3.130 $\pm$ 0.097 && XZ Car & 473 $\pm$ 18 & -7.277 $\pm$ 0.019 & 2.622 $\pm$ 0.017&  15.3 & 0  \\ 
        Cl X Pup & $113.353  $ & $-20.481$  & 125 & 363 $\pm$ 6 & -2.166  $\pm$ 0.149 & 3.125 $\pm$ 0.205 && X Pup & 397 $\pm$ 20 & -1.236 $\pm$ 0.016 & 1.786 $\pm$ 0.018&  24.4 & 0 \\ 
\bottomrule
\end{tabular}
\caption{Silver sample of cluster Cepheids.}
 \label{tab:Silver}

\vspace{2\baselineskip}
\begin{tabular}{lrrrrrrclrrrrr}\toprule
 \multicolumn{5}{c}{Cluster parameters} & \phantom{   }& \multicolumn{5}{c}{Cepheid parameters}\\
\cmidrule(lr){0-6} \cmidrule(lr){9-14} 
Cluster& $\alpha (^{\circ})$  & $\delta (^{\circ})$& N  & $\varpi$ ($\mu$as) & $\mu^{*}_{\alpha}$ (mas/yr)  & $\mu_{\delta}$ (mas/yr)  && Cepheid& $\varpi$ ($\mu$as) & $\mu^{*}_{\alpha}$ (mas/yr)  & $\mu_{\delta}$ (mas/yr) & Sep   (pc) & M.P \\ \midrule
Asterism BB Cen & $178.358  $ & $-62.608 $& 284 & 316 $\pm$ 6 & -6.196 $\pm$ 0.378 & 0.781 $\pm$ 0.173 && BB Cen &  330 $\pm$  11 & -6.43 $\pm$ 0.011 & 0.978 $\pm$ 0.013 & 15.0 &   0.83 \\ 
Asterism V620 Pup & $119.732  $ & $ -29.490 $& 118 & 293 $\pm$ 7 & -2.346 $\pm$ 0.232 & 2.895 $\pm$ 0.350 && V620 Pup &  268 $\pm$  14 & -2.282 $\pm$ 0.012 & 3.668 $\pm$ 0.013 & 17.3& 0.45 \\ 
\bottomrule
\end{tabular}
\caption{Bronze sample of cluster Cepheids.}
\label{tab:Copper}

\vspace{2\baselineskip}
\begin{tabular}{llll}\toprule
Cepheid  &   Cluster  & Reason of rejection & Reference \\ \midrule
BB Sgr & Collinder 394   &   $7 \sigma$ difference in parallax  & \cite{gieren1997very,usenko2019spectroscopic}  \\
OGLE-GD-CEP-1175 &   NGC 6193  &        Position in the color magnitude diagram &   \cite{medina2021revisited}  \\
WISE J124231.0-625132 &  NGC 4609  &  Position in the color magnitude diagram &  \cite{medina2021revisited} \\  
SX Car & ASCC 61  &   No host cluster detected at this location &  \cite{anderson2013cepheids,chen2015search}   \\
S Mus   &  ASCC 69    &  No host cluster detected at this location &  \cite{anderson2013cepheids,chen2015search}   \\    
\bottomrule
\end{tabular}
\caption{Cepheids considered as possible cluster members in the literature that were  not found to be bona fide cluster Cepheids here.}
\label{tab:Rejected}
\end{sidewaystable*}

\subsection{Bronze sample\label{sec:copper}}
Clusters reported as part of the  Gold and Silver samples can be clearly distinguished from field stars in position and proper motion. However, these distinctions were less clear in the case of possible host clusters (tentatively labeled asterisms) reported here as part of the Bronze sample. Additionally, the \gaia\ CMDs exhibit two main sequences, suggesting likely fore- or background contamination, perhaps by spiral arms being crossed (cf. Figure~\ref{fig:examples2}). Unfortunately, the cluster membership probabilities provided by HDBSCAN do not allow us to filter out contaminants. However, there appear to be a clear overdensities in parallax space for stars in the vicinity of both BB~Cen and V0620~Pup, and we note that the computed likelihoods for the Cepheid are  high and fully consistent  with cluster membership, assuming the cluster is real.

\begin{figure*}[!htp]
    \centering
    \includegraphics[scale=0.52]{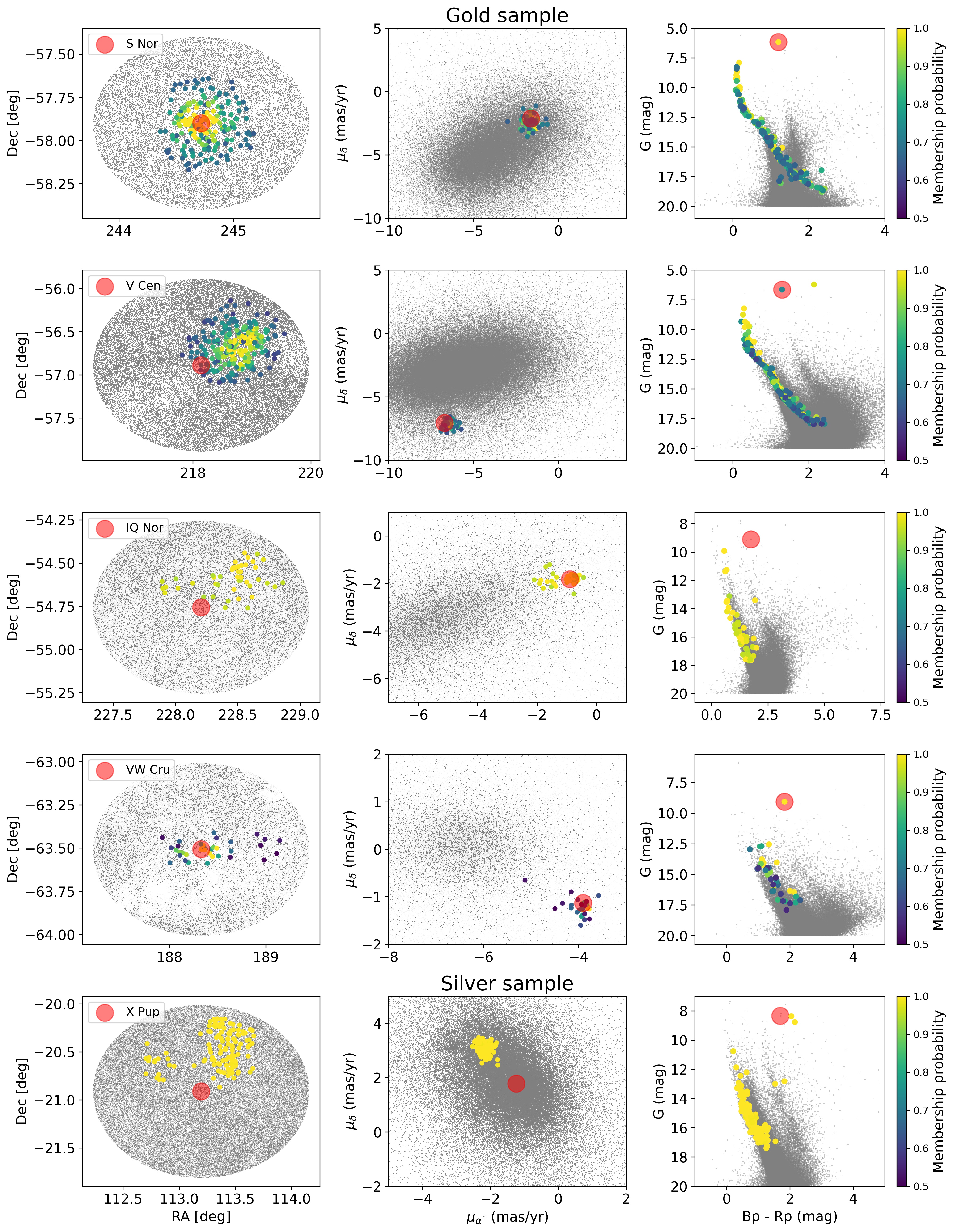}
    \caption{Position in the sky, position in the proper motion space, and color magnitude diagram for different cluster Cepheids. Background stars are shown in gray, and the cluster  membership probability is color-coded. Light colors indicate high probability. Cepheids are shown as labeled using large filled red circles. Cepheids detected as cluster members by HDBSCAN also feature an overplotted symbol to illustrate membership probability.}
    \label{fig:examples1}
\end{figure*}

\begin{figure*}[!htp]
    \centering
    \includegraphics[scale=0.52]{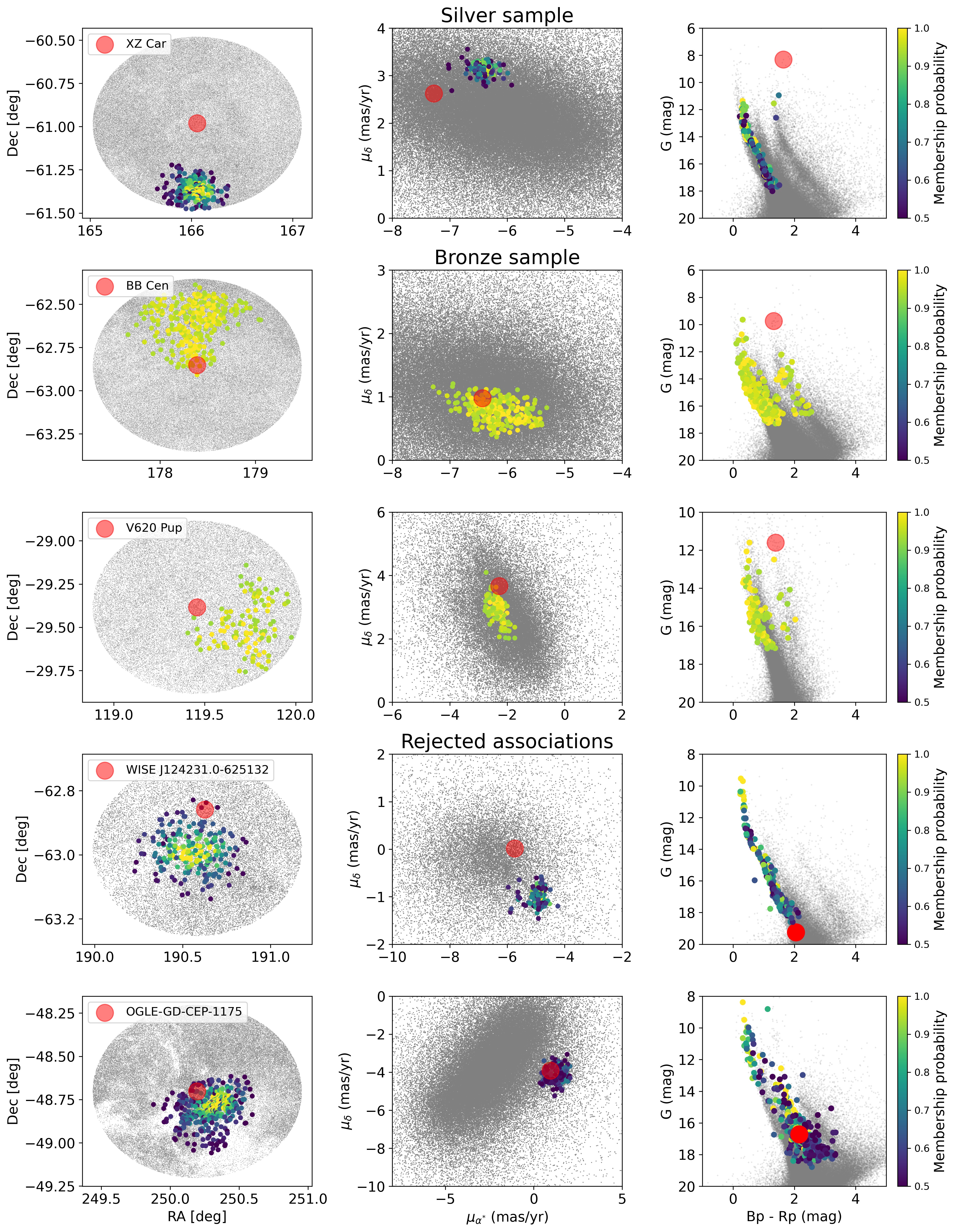}
    \caption{Graphical representation of membership constraints for specific examples of cluster Cepheids in the Silver (XZ~Car) and Bronze (BB~Cen, V620~Pup) samples, as well as two rejected associations.} 
 \label{fig:examples2}
\end{figure*}

\begin{figure*}[!htp]
    \centering
    \includegraphics[scale=0.5]{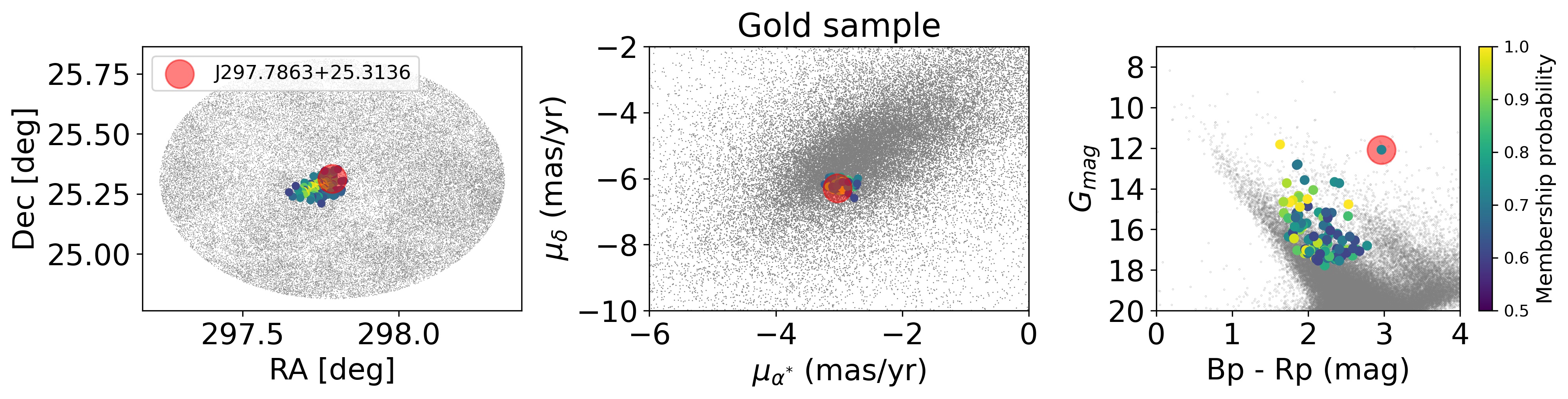}
    \caption{Cluster Czernik 41 and the  Cepheid ATO J297.7863+25.3136. The cluster is  a clear overdensity on the sky and in the proper motion space. However, the CMD does not exhibit a clean main sequence with member stars at atypically red colors, indicating high extinction, which is also reflected by varying levels of background stars amid the gray points on the left. } 
 \label{fig:ATLAS}
\end{figure*}

\subsection{Rejected associations} \label{sec:rejected}

Our analysis refuted the cluster membership of several Cepheids previously considered as cluster members in the literature, and these cases are listed in Table\,\ref{tab:Rejected}. Specifically, OGLE-GD-CEP-1175 and WISE J124231.0-625132 were previously reported as members of NGC 6193 and NGC 4609 \citep{medina2021revisited}, respectively. They are both  too faint to be among the most luminous evolved stars in the recovered host clusters, however (cf. Fig. ~\ref{fig:examples2}). BB~Sgr was previously considered to be a member of Collinder~394 \citep{gieren1997very,usenko2019spectroscopic}. However, \cite{medina2021revisited} found that neither parallax nor proper motion match for this combination. Notably, the parallax differs by $7\sigma$, which reliably excludes membership.

Data from the All-Sky Compiled Catalogue \citep{kharchenko2001all} led \citet{anderson2013cepheids} and \citet{chen2015search} to consider the Cepheids SX~Car and S~Mus potential cluster members of the clusters ASCC~61 and 69, respectively. \citet{Evans2014SMus} reported the detection of a population of X-ray sources near the position of S~Mus and interpreted this population as evidence for young stars in the cluster ASCC~69. At the reported cluster distances of $1700$ and $1000$\,pc, \gaia\ data should allow the detection of even relatively small clusters near the positions of the two Cepheids. However, because no clusters are detected, we conclude that S~Mus and SX~Car are not bona fide cluster Cepheids.

\section{LL and \gaia\ zeropoint offset calibration}\label{sec:PLR}

In this section, we calibrate period luminosity relations for MW Cepheids that pulsate in the fundamental mode while simultaneously investigating residual parallax offsets that are applicable after applying the L21 corrections. Section\,\ref{sec:LLdata} describes the observational data for MW Cepheids, Sect.\,\ref{sec:LMC} contains a cross-check of the expected zero residual offset applicable to cluster parallaxes using the LMC, and Sect.\,\ref{sec:combined} describes the calibration of the MW LL using combined cluster and field Cepheids. 

\subsection{Milky Way Cepheids\label{sec:LLdata}\label{sec:MW}}

\begin{table*}{}
 \centering
\caption{Astrometric and photometric constraints applied to the MW Cepheid sample.\label{tab:constraints}}
\begin{tabular}{ccccc}\toprule
\multicolumn{1}{c}{Astrometric constraints} &  \multirow{1}{*}{Photometric constraints}   \\
 \cmidrule(lr){1-1}\cmidrule(lr){2-2} 
RUWE $<1.4$     &   $0.8 < Bp - Rp < 2.75$ \\ 
$ \varpi/\sigma_\varpi \geq 8.5$ &  num\_clean\_epochs\_g $ > 15$  \\ 
astrometric\_chi2\_al $< 3000$ &   num\_clean\_epochs\_bp $ > 15$  \\ 
astrometric\_excess\_noise $<0.25$  & num\_clean\_epochs\_rp $ > 15$
   \\ 
astrometric\_excess\_noise\_sig $<70$ & ipd\_frac\_multi\_peak$<7$    \\ 
 &  int\_average\_g$>6$  \\ 
\bottomrule
\end{tabular}
\tablefoot{The constraints relate to parameters given in \gaia\ DR3 data tables \texttt{gaia\_source} and \texttt{vari\_cepheid}. Astrometric constraints\footnote{Descriptions available here: \url{https://gea.esac.esa.int/archive/documentation/GDR3/Gaia_archive/chap_datamodel/sec_dm_main_source_catalogue/ssec_dm_gaia_source.html}} are applied to all Cepheids used in this work and reproduce the sample of Cepheids used by \citet{riess2021cosmic}. \texttt{astrometric\_chi2\_al} quantifies the goodness of fit in the along-scan direction without taking into account \texttt{astrometric\_excess\_noise}. Positive values of \texttt{astrometric\_excess\_noise} indicate that the source may not be astrometrically well behaved, and this excess noise may be relevant if \texttt{astrometric\_excess\_noise\_sig} $> 2$. Since currently, a detailed guidance for how to use these parameters is lacking, we adopted very conservative cuts to remove the clearest outliers. Photometric constraints are applied only to the sample of Cepheids for which \gaia\ photometry is used. In particular, the parameter \texttt{ipd\_frac\_multi\_peak} specifies the percentage of multiply peaked \gaia\ windows that were accepted by the image parameter determination. We adopted a constraint in this parameter to avoid blending of the Cepheid photometry with nearby sources, which particularly applies to the $Bp$ and $Rp$ spectrophotometry. An overview of these samples is given in Table\,\ref{tab:GaiaSample}.}
\end{table*}

\begin{table*}[]
    \centering
    \caption{MW Cepheid samples used to calibrate the Galactic Cepheid LL in various passbands}
    \begin{tabular}{@{}lllccccccccc}\toprule
         Cepheid & Gaia DR3 source id & $P$ & $\varpi$ & $\varpi_{\rm corr}$ & $W_H$ & $W_G$  & $Bp$ & $V$ & $G$ & $Rp$  & $F160W$ \\
          & & (d) & (mas) & (mas) & \\ \midrule
          AA Gem & 3430067092837622272 & 11.297 & 0.274 & 0.311 & $W_H$ &  & & &  &  & $F160W$\\
          AA Mon & 3102535635624415872 & 3.937  & 0.313 & 0.316 &  &   & $Bp$ & $V$ & $G$ & $Rp$   & \\
          AB Cam & 473239154746762112 & 5.788 & 0.212 & 0.241 &  & &   $Bp$ & $V$ & $G$ & $Rp$  & \\
          \ldots & \ldots & \ldots & \ldots & \ldots & \ldots & \ldots & \ldots & \ldots & \ldots & \ldots & \ldots \\
          
         \bottomrule
    \end{tabular}
    \tablefoot{The complete version of this table is available at the CDS. $\varpi$ is the Cepheid parallax as obtained from \gaia\ DR3, and $\varpi_{\rm corr}$ lists the parallax corrected for the L21 offset.}
    \label{tab:GaiaSample}
\end{table*}

We compiled samples of fundamental-mode MW Cepheids based on the astrometric and photometric quality criteria tabulated in Table\,\ref{tab:constraints}. The astrometric constraints were compiled such as to reproduce the sample of 68 low-reddening MW Cepheids observed by the SH0ES team using \hst\ \citep{riess2018standards,riess2021cosmic}. However, a larger sample of Cepheids is considered in other photometric bands and using \gaia\ photometry. Hence, we added cuts based on astrometric goodness-of-fit parameters to remove Cepheids whose astrometry was very likely flawed, such as RX~Cam, the only Cepheid for which an orbital parallax solution is available in \gaia\ DR3. The photometric criteria we adopted include a magnitude cut to avoid saturated stars, a color cut to limit exposure to reddening, a cut on the number of available photometric epochs based on which the mean magnitudes were computed, and the parameter \texttt{ipd\_frac\_multi\_peak} $< 7,$ which was adopted to limit exposure to blended sources. We further adopted a cut on period for the \gaia\ sample $P > 3.9$d to avoid exposure to misclassified overtone Cepheids. The most stringent cut in practice is that we require individual iron abundance measurements based on high-resolution spectroscopy for all sample stars \citep{genovali2014fine,genovali2015alpha}. The final sample of fundamental-mode classical Cepheids for $W_G$  contains $225$ stars and is listed in Table\,\ref{tab:GaiaSample}.  

We compiled ground-based photometry in the  Johnson $V$ and Cousins $I $ bands from \citet{groenewegen2018cepheid} and \citet{breuval2020milky,breuval2021influence}. This dataset has been homogenized by \citet{groenewegen2018cepheid} and was studied extensively. It mainly includes $V-$ and $I-$band data reported by \citet{melnik2015phot}, which are based on observations by L.~Berdnikov \citep[e.g.,][]{berdnikov2008vizier}. 
Reddening values, $\mathrm{E(B - V)}$ for Galactic Cepheids are taken from \citet{Fernie} and scaled by a factor of $0.94$ following \citet{groenewegen2018cepheid}. We also computed reddening-free Wesenheit magnitudes \citep{Madore1982} using $V$ and $I-$band data, $W_{VI}$ (cf. below). 

We collected \gaia\ DR3 photometry in \gaia\ $G $ band, as well as integrated $Bp$ and $Rp$ spectrophotometry  \citep{gaiadr3.cepheid,gaiaedr3.phot}. Specifically, we used intensity-averaged magnitudes from \gaia\ CU7 Specific objects studies (parameters \texttt{int\_average\_g}, \texttt{int\_average\_g\_error} and analogous for $Bp$ and $Rp$ from table \texttt{gaiadr3.vari\_cepheid}) published as part of the Gaia DR3 variability analysis for Cepheids \citep{gaiadr3.cepheid,gaiadr3.vari}. We also computed reddening-free Wesenheit magnitudes, $W_G$, based on $G$, $Bp$, and $Rp$ as stated below.

Finally, we collected \hst\ WFC3-IR F160W photometry for MW Cepheids from \citet{riess2019large} and \citet{2022arXiv220606212R}, as well as their reported reddening-free NIR Wesenheit magnitudes $W_H$. Benefits of this homogeneous \hst\ dataset include the excellent calibration of the \hst\ photometric system, homogeneity with respect to extragalactic Cepheids, high spatial resolution, and the lack of time- and location-specific calibration issues typical of ground-based NIR photometry. We also experimented with ground-based near-IR photometry available from a range of literature references following  \citet{breuval2021influence}, notably combining ground-based $J,H,K_s$ photometry from \citet{laney1992jhkl}, \citet{monson2011vizier}, and  \citet{genovali2014fine}. However, the homogenization of these data sets is not as straightforward due to different photometric systems in use (e.g.,  improvements in detector technology), the calibration of atmospheric absorption in the NIR, and the standardization of NIR passbands. After some tests, and notably in comparison with the \hst\ F160W photometry available from \citet{riess2019large}, we  discarded ground-based NIR photometry as not sufficiently accurate for the purposes of our study.

\hst\ WFC3-IR observations are subject to count-rate nonlinearity (CRNL) at the level of $0.0075 \pm 0.006$\,mag/dex \citep{riessCRNL}. We took these CRNL corrections into acount when we compared Cepheid samples spanning a significant dynamic range, that is, when we compared MW Cepheids to extragalactic Cepheids, such as those in the LMC, or when we compared them to Cepheids in supernova-host galaxies (SN-hosts). CRNL corrections to offset differences among MW Cepheids alone are at the level of $1-2$\,mmag and were therefore neglected.

We used the following definitions for Wesenheit magnitudes $W_{VI}$ \citep{breuval2022improved}, $W_{H}$ \citep{riess2016H0}, and $W_G$ \citep{ripepi2018re}: 

\begin{equation}
    W_{VI} = I - 1.239\cdot(V-I)\ , \label{eq:WVI}
\end{equation}
\begin{equation}
    W_{H} =  F160W - 0.386 \cdot (F555W - F814W)\ , \mathrm{ and} \label{eq:WH}
\end{equation}
\begin{equation}
    W_{G} = G - 1.921\cdot(Bp - Rp) \ .  \label{eq:WG}
\end{equation}

Extinction corrections were applied using reddening coefficients calculated for a \citet{fitzpatrick1999correcting} reddening law with $R_V=3.3$ and a spectral energy distribution representative of a $10$\,d Cepheid near the center of the instability strip \citep[cf.][]{2022A&A...658A.148A} as given by a \citet{castelli2004new}  model atmosphere with $T_{\rm eff} =5400 K$, [Fe/H]$=0.0$, $\log{g}=1.5$. Specifically, this yields  $R_{V_{\text{Johnson}}} = 3.553 $, $R_{I_{\text{Cousins}}} = 2.095 $,  $R_{F160W} = 0.674$ and $R_{Bp} = 3.701$, $R_{G} = 2.991$, $R_{Rp} = 2.196$, where the subscript F160W refers to the \hst\ WFC3-IR system. All filter profiles were downloaded from the Spanish VO filter profile service\footnote{\url{http://svo2.cab.inta-csic.es/theory/fps/}}. These values were used in conjunction with color excess values defined for Johnson-Cousins $E(B-V)$ to estimate extinction in the respective bands. We also compiled individual iron abundances from the literature ensuring a common solar iron abundance (cf. Sec.\,\ref{sec:ZP}).

\subsection{Confirming the adequacy of L21 parallax corrections for cluster parallaxes using the LMC}\label{sec:ZP}\label{sec:LMC}

\cite{lindegren2021gaia2} provided a recipe for correcting systematic parallax errors related to source magnitude, color, and sky-position (ecliptic latitude) based on millions of quasars and LMC stars as well as 7000 bright physical stellar pairs. However, previous articles have presented evidence that residual parallax offsets need to be applied even after the L21 corrections are applied. For example, \citet{riess2021cosmic} determined an additional constant parallax offset of $  14 \pm 6$ $ \mu$as based on 75 Galactic Cepheids  in the magnitude range $6<G<12$, and these residual offsets are now well documented using different methods and stellar types \citep[e.g.,][]{zinn2019confirmation,zinn2021,khan2019new,schoenrich2019,stassun2021,ren2021EBplx,wang2022lamost,flynn2022clusters}. Hence, an accurate LL calibration based on Cepheid parallaxes requires solving for the residual offset applicable to the sample of stars used in the calibration.

However, recent work based on open and globular clusters as well as the Magellanic Clouds has shown that the L21 recipe accurately corrects parallax systematics of stars fainter than $G > 13$\,mag \citep{flynn2022clusters,apellaniz2022clustersclouds}. As a result, average cluster parallaxes based on L21-corrected member stars in this magnitude range are particularly useful for LL calibration because no further offsets need to be determined, that is, $\Delta \varpi_{\mathrm{Cl}} = 0$. Average cluster parallaxes can therefore inform the residual parallax offset applicable to Cepheid parallaxes, $\Delta \varpi_{\mathrm{Cep}}$. This is done in Sect.\,\ref{sec:combined}. However, prior to adopting $\Delta \varpi_{\mathrm{Cl}} = 0$, we decided to verify the validity of this approach using observations of Cepheids in the LMC, whose distance $\mu_{\text{DEBs}}$ is known with an accuracy of $1.3\%$ from detached eclipsing binary stars \citep{pietrzynski2019distance}.

We compiled Johnson-Cousins $V-$ and $I-$band photometry of LMC Cepheids from the OGLE-III catalog of variable stars \citep{soszynski2017ogle} and selected fundamental-mode OGLE-III Cepheids within the matching period range of cluster Cepheids ($3.9-45$\,d) and cross-matched their positions (maximum search radius  $2''$) with \gaia\ DR3 positions to obtain \gaia\ $G-$band, $Bp$, and $Rp$  photometry from the SOS Cepheid list \citep[\texttt{gaiadr3.vari\_cepheid}]{gaiadr3.cepheid}. The accuracy of the cross-match was verified by considering the agreement between periods reported by OGLE and \gaia. We adopted the OGLE-III Cepheid sample instead of the \gaia\ DR3 list of Cepheids in the LMC direction because a) geometric corrections (cf. below) are well described for this sky region \citep{pietrzynski2019distance}, and b) the classification of Cepheids in OGLE-III benefits from longer time series and long-standing human experience in classification. Since OGLE-III covers the main part of the LMC disk, and thus the bulk of the Cepheid population, including outer regions from \gaia\ is not expected to add a significantly greater number of Cepheids outweighing the downsides related to the geometric correction. We used reddening maps based on red clump stars \citep{2021ApJS..252...23S} to correct for extinction using the values of $R_\lambda$ mentioned in Sect.\,\ref{sec:MW} and the conversion $E(V-I) = 0.686 \cdot E(B-V)$ derived analogously.

For our NIR analysis, we used \hst\ WFC3 observations of 70 LMC Cepheids \citep{riess2019large} because they can be directly compared to the \hst\ observations of MW cluster Cepheids  \citep{2022arXiv220801045R} after the appropriate CRNL corrections are applied. Because the NIR Wesenheit magnitudes, $W_{H}^{HST}$, reported by \citet{riess2019large} already include a CRNL correction applicable  for the comparison with Cepheids in the SN-host sample, we recomputed $W_H$ using Eq.\,\ref{eq:WH} and their original \hst\ observations in the individual passbands $F555W$, $F814W,$ and $F160W$. We then applied appropriate CRNL corrections (average of $0.010$\,mag) to account for the flux difference of $0.9-1.8\,$dex between MW cluster and LMC Cepheids.

We applied geometric corrections to apparent magnitudes following \citet{jacyszyn2016ogle}, effectively treating all LMC Cepheids as though they were observed at the same distance, determined to an accuracy of $1.3\%$ using detached eclipsing binary systems \citep{pietrzynski2019distance}. As a result of this correction, the effect of the LMC intrinsic depth on the observed scatter in the LL is minimized. This is necessary due to the large sky region covered by OGLE-III ($1.7$\,kpc) and to ensure that the distance estimate to the LMC reflects the distance to the barycenter of the detached eclipsing binaries. Moreover, the correction decreases the observed scatter in the LL, resulting in a slight ($\sim0.004\,$mag) improvement in the uncertainties for the LL intercept $\beta$.

For LMC Cepheids, we fit linear LLs of the form $m =  \alpha (\log P - \log P_{0} ) + \beta^{'}$ using a least-squares fitting procedure and a $2.7 \sigma$ outlier rejection (applying Chauvenet’s criterion for the {\it HST} LMC Cepheid sample);  $m$ denotes apparent magnitudes corrected to the LMC barycenter. Depending on the photometric data set, the samples used in the fit contained between $68$ and $712$ LMC Cepheids. The results for a range of individual photometric bands and Wesenheit magnitudes are listed in Table\,\ref{tab:ZPC}, including the number of available Cepheids, and the assumed intrinsic width of the LL.

\begin{table*}{}
 \centering
\caption{LLs of the form $m = \alpha(\log{P - 10}) + \beta $ fit to LMC Cepheids\label{tab:ZPC}} 
\begin{tabular}{ccccc}\toprule
\multirow{1}{*}{Band} & \multicolumn{3}{c}{LMC} &  \multirow{1}{*}{WIS}   \\
 \cmidrule(lr){1-1}\cmidrule(lr){2-4} \cmidrule(lr){5-5}
 & $\beta$     & $\alpha$ &  $n_{\text{LMC}}$  & \\
 & (mag) & (mag/$\log{P}$) & & (mag) \\ \midrule
$Bp$ & $14.329  \pm 0.016 $ &  $ -2.717 \pm 0.055$ & 546 & 0.23 \\
$V$ & $ 14.187  \pm0.014 $ &  $-2.669 \pm 0.049$ & 701& 0.22 \\
$G$ & $  14.066\pm0.013  $ &  $ -2.857 \pm 0.046$ & 546 & 0.19 \\
$Rp$ & $ 13.611  \pm 0.011$ &  $ -2.962 \pm 0.037$ & 546 & 0.16 \\
$I$ & $ 13.580 \pm 0.011 $ &  $ -2.872 \pm 0.038$ & 712 & 0.14\\
$F160W^{a}$ & $ 12.923  \pm  0.014$ &  $ -3.229  \pm 0.056$ & 68 & 0.09\\
$W_{VI}$  & $  12.870 \pm 0.008$ &  $ -3.181 \pm 0.029$& 684 & 0.08\\
$W_{G}$& $ 12.669 \pm 0.005 $  & $  -3.347 \pm   0.020 $ &  543 & 0.10\\
$W_{H}^{a}$  & $ 12.643 \pm 0.011  $ &  $  -3.290 \pm 0.044 $ & 68 & 0.06\\
\bottomrule
\end{tabular}
\tablefoot{ $\beta$ is expressed here in apparent magnitudes. The last column indicates the intrinsic width of the LL due to the finite width of the instability strip (WIS), adopted from \citet{breuval2022improved}.  Superscript $^a$ indicates that no \hst\ F160W-IR CRNL corrections were applied to  observations of LMC Cepheids for this comparison. Magnitudes in the NIR Wesenheit function were recomputed using Eq.\,\ref{eq:WH} based on the observations reported by \citet{riess2019large}.}
\end{table*}

To determine the validity of the expected $\Delta \varpi_{\mathrm{Cl}}=0$, we computed the absolute magnitude of LMC Cepheids by applying the distance modulus measured obtained using detached eclipsing binaries \citep[$\mu_{\mathrm{DEB}}=18.477 \pm 0.004 \, (\mathrm{stat}) \pm 0.026 \, (\mathrm{syst}) $\,mag]{pietrzynski2019distance},
\begin{align}
\text{M} &= \alpha (\log{P - 10\ \mathrm{[d]}}) + \beta^{'} - \mu_{\mathrm{DEB}} \\
         &= \alpha (\log{P - 10\ \mathrm{[d]}}) + \delta^{'} - \mu_{\mathrm{DEB}}  + \gamma \Delta_{\mathrm{[Fe/H]}} \label{eq:LMCLL} \ .
\end{align}
These absolute magnitudes of LMC metallicity Cepheids were then compared to MW Cepheids using the astrometry-based luminosity \citep[ABL]{arenou1998distances}, which avoids the issue of inverting parallaxes to obtain distances,
\begin{align}
\text{ABL} &=  10^{\frac{M}{5}} = (\varpi+\Delta \varpi_{\mathrm{Cl}}) 10^{\frac{m -10 }{5}}\ .\label{eq:ABL2}
\end{align}
Superscript $^{'}$ in Eq.\,\ref{eq:LMCLL} implies that $\beta$ and $\delta$ are given in apparent magnitudes after applying geometric corrections to the LMC Cepheids. $\beta$ denotes the LL intercept at the average sample metallicity, $\delta = \beta - \gamma \mathrm{[Fe/H]}$ is the LL intercept corrected to solar metallicity, and $\Delta_{\mathrm{[Fe/H]}}$ is the difference in iron abundance between the MW and LMC Cepheid samples. Table\,\ref{tab:ZPC2}  lists the results of this comparison for six individual photometric bands and three Wesenheit formulations.

The metallicity difference between LMC and MW Cepheid requires careful consideration. For the LMC, we adopted a common mean iron abundance, $\mathrm{[Fe/H]_{LMC}} = -0.409 \pm 0.003$, based on the recently remeasured average iron abundances of LMC Cepheids that has been shown to be consistent with a single value \citep[dispersion $0.076$\,dex]{romaniello2022iron}. For MW cluster Cepheids, we adopted individual iron abundances as described above and compiled in Table\,\ref{tab:metal}. Although several improvements in the determination of $\gamma$ have been recently presented \citep{gieren2018effect,breuval2021influence,breuval2022improved,2022A&A...659A.167R}, we here preferred to use $\gamma$ as a free parameter, while first fixing $\Delta\varpi_{\mathrm{Cl}}=0$ and then performing the same comparison while fitting for $\gamma$ and $\Delta\varpi_{\mathrm{Cl}}$ simultaneously. The individual MW Cepheid abundances are compiled in Table\,\ref{tab:metal}.

\begin{table*}{}
 \centering
\caption{Metallicity term $\gamma$  and zeropoint offset $\Delta \varpi$  obtained by comparing Gold sample cluster Cepheids to the LMC LL using Equation (\ref{eq:ABL2}).\label{tab:ZPC2}} 
\begin{tabular}{cccccccc}\toprule
\multirow{1}{*}{Band} & \multicolumn{2}{c}{$\gamma$ free and $\Delta \varpi = 0$ }   & \multicolumn{2}{c}{$\gamma$ and $\Delta \varpi$ free}    \\
 \cmidrule(lr){1-1}\cmidrule(lr){2-3} \cmidrule(lr){4-5} \cmidrule(lr){6-7}
        & $\gamma$     & $\Delta \varpi$ & $\gamma$     & $\Delta \varpi$  &  $n_{\text{MW}}$  &\\ 
        & (mag/dex) & ($\mu$as) & (mag/dex) & ($\mu$as) & & \\ \midrule
$Bp$ &  $ -0.041  \pm 0.098 $ &$  0 $& $ -0.191  \pm 0.255 $ & $-18 \pm  27 $  &  23\\ 
$V$ & $ 0.015  \pm 0.090$ &$ 0 $  & $ 0.029 \pm 0.216$ & $ 2 \pm 24 $ &22\\ 
$G$ &  $ -0.300  \pm 0.102  $ & $  0 $  & $ -0.447  \pm 0.265 $ & $ -18  \pm 28$  & 23\\ 
$Rp$ & $ -0.168 \pm 0.075  $ &$ 0 $  & $ -0.187 \pm  0.183 $ & $ -2 \pm  21 $   & 23\\ 
$I$ &  $ -0.083 \pm 0.102$ &$  0 $ & $ -0.150 \pm 0.217$ & $ -8 \pm  23 $   &15\\ 
$F160W^a$ &  $ -0.259 \pm 0.062 $ &$  0 $ & $ -0.216\pm 0.142  $ &$  5 \pm  15 $  &  15\\ 
$W_{VI}$ & $  -0.035 \pm 0.098 $ &$ 0 $  & $ -0.146 \pm  0.206$ &$ -14 \pm  22$  &  15\\ 
$W_{G}$ &  $  -0.394 \pm 0.058 $ &$ 0$   & $ -0.448  \pm 0.141 $ &$ -6 \pm 15 $    &  26\\
$W_{H}^a$ &  $ -0.287 \pm 0.059 $ &$ 0  $  & $ -0.297 \pm 0.134  $ &$ -1 \pm14  $   &  15\\ 
\bottomrule
\end{tabular}
\tablefoot{Wesenheit magnitudes of cluster Cepheids, $W_{H}$, were computed using Eq.\,\ref{eq:WH} using the photometric data for the individual passbands presented by \citet{2022arXiv220801045R}. Cluster parallaxes were bias corrected using the L21 approach. Superscript $^a$ indicates that \hst\ WFC-IR CRNL corrections have been applied to account for the $0.9$ to $1.8$ dex difference in flux among MW cluster Cepheids and LMC Cepheids (mean correction $0.010$\,mag). The weighted mean and associated uncertainty for $\Delta \varpi_{\mathrm{Cl}}$ is  $-4 \pm 6\,\mu$as.} 
\end{table*}

Our results for $\gamma$ listed in Table\,\ref{tab:ZPC2} show that metal-rich Cepheids are typically brighter than metal poor Cepheids in each of the photometric bands as well as the three Wesenheit formulations. This echoes recent results by \citet{breuval2022improved}, albeit at lower precision because the metallicity range we considered is limited. Additionally, as noted by Breuval et al., our results are consistent with predictions of $\gamma$ derived from Geneva stellar evolution models \citep{anderson2016effect}. We further confirm the particularly strong metallicity dependence in \gaia\ $G $ band and the \gaia\ Wesenheit function $W_G$ reported by \citet{breuval2022improved} and \citet{2022A&A...659A.167R}, while neither $Bp$ nor $Rp$ exhibit such a steep trend with metallicity.

Concerning $\Delta \varpi_{\mathrm{Cl}}$, we find residual offsets consistent with $0$ to within $1\sigma$ in all nine cases, and a weighted mean value of $\Delta \varpi = - 4 \pm 6\,\mu$as. Additionally, we note that the values of $\gamma$ obtained when fixing $\Delta \varpi_{\mathrm{Cl}}=0$ are consistent to within their uncertainties with $\gamma$ values obtained when both parameters are free, as well as with recent literature results. In summary, our comparison involving the LMC thus strongly supports that the average cluster parallaxes determined above exhibit no evidence of residual parallax offsets beyond the L21 corrections.

\begingroup
\setlength{\tabcolsep}{3.pt} 
\begin{table*}{}
  \centering
\caption{Information used for determining the Galactic LL using cluster Cepheids.}
\begin{tabular}{ccccccccccc}\toprule
Name & Period & $\varpi$  & $\sigma_{\text{stat}}$ &  $\sigma_{\text{cov}}$ &  $\sigma_{\text{total}}$ & $W_{G}$  & $W_{H}^a$ & [Fe/H]  & Reference & E(B-V)  \\ 
 & (d) & ($\mu$as) & ($\mu$as) & ($\mu$as) & ($\mu$as) & (mag) & (mag) & & & (mag) \\ \midrule
CE Cas A & 5.141 & $322.5 $ &  3.1 & 6.7 & 7.4 & $ 7.628 \pm 0.073 $ &   & ~ & \\  
CE Cas B & 4.479 & $322.5$ &  3.1 & 6.7 & 7.4 & $ 7.759 \pm 0.081 $ &   & ~ & \\  
CF Cas & 4.875 & $322.5 $&  3.1 & 6.7 & 7.4 &  $ 7.642 \pm 0.010 $  & $ 7.641 \pm 0.030$ &  $0.02 \pm 0.06$ &LL11 & 0.556 $\pm$ 0.021 \\  
CG Cas & 4.366 & $335.7 $ &  3.0 & 6.7 & 7.3 & $ 7.687  \pm 0.030 $ & $ 7.652 \pm 0.034$ & $0.09 \pm  0.06$ & LL11  & 0.667 $\pm$ 0.009\\
CM Sct & 3.917 & $442.9$ &  2.2 & 6.2 & 6.6 &  $ 7.046 \pm      0.014 $ &                    & $ 0.15 \pm 0.06 $ &  LL11 & 0.775 $\pm$ 0.045\\  
CS Vel & 5.905 & $281.2$ &  2.8 & 6.7 & 7.3 &  $ 7.676 \pm 0.006 $  & $7.669 \pm 0.029$ & $0.12 \pm 0.06$ & G14 & $0.716 \pm 0.027$ \\  
CV Mon & 5.379 & $585.2$ &  6.9 & 6.6 & 9.6 &  $ 6.198 \pm 0.006 $  & $6.205  \pm 0.029$ & $0.09\pm 0.09$ & G14& 0.705 $\pm$  0.018  \\ 
DL Cas & 8.001 & $556.6$ &  2.2 & 6.5 & 6.9 &  $ 5.597 \pm 0.006 $  & $5.645  \pm  0.021$ &  $ -0.01 \pm 0.08$  &  L11&  $0.487 \pm 0.005$\\          
EV Sct$^{(*)}$ & 3.090 & $ 504.0$ &  2.8 & 6.4 & 7.0 & $ 6.541 \pm 0.010 $ &             & $  0.09  \pm 0.07  $  & G15& \\ 
GH Lup & 9.276 & $877.8$ & 6.1  & 6.4 & 8.8 &  $4.448 \pm 0.008$& & $0.13\pm 0.06$ &  G14 &   $0.328 \pm 0.011$ \\
IQ Nor & 8.220 & $543.6 $ &  7.2 & 6.1 & 9.4 & $  5.711 \pm 0.016 $ &                    & $0.22 \pm 0.07$ & G15 & $ 0.676 \pm 0.044$\\
NO Cas$^{(*)}$ & 2.582 & $ 317.0$ & 2.7 & 6.4 & 7.0  & $ 8.113 \pm 0.018 $ &             & & & \\ 
QZ Nor$^{(*)}$ & 3.786 & $ 513.4  $ & 1.4 & 6.5& 6.6 & $ 6.379 \pm 0.006 $ &             & $  0.21 \pm 0.06 $&   G15& $0.289 \pm 0.020$\\ 
RS Ori & 7.567 & $609.6$ & 3.5 & 6.3& 7.2 & $ 5.571     \pm 0.036$ & $ 5.603 \pm 0.030$      & $0.11 \pm 0.09$&  G15 & $0.332        \pm 0.010$\\  
S Nor & 9.754 & $1073.2$ & 3.8 & 6.1& 7.2 & $ 3.932 \pm 0.014$ & $ 3.940  \pm 0.014$     & $0.02 \pm 0.09$ & R08 & $0.182 \pm 0.008$\\
ST Tau & 4.034 &  $953.0$ & 5.6 & 6.0 & 8.3& $ 5.608 \pm 0.019$&                         & $ -0.14 \pm 0.15$ &  L07 & $0.328 \pm 0.006$ \\ 
SV Vul & 44.993 & $424.8$ & 5.9 & 6.6 & 8.9 & $ 3.393  \pm 0.014 $ & $3.589 \pm 0.035$   & $0.05\pm 0.08$ & L11& $0.474 \pm 0.024$ \\  
SX Vel & 9.550 & $497.0 $ & 4.0 & 6.0 & 7.2 & $ 5.746 \pm 0.019 $  &                     &  $-0.18\pm 0.07$ & L07& $0.237 \pm 0.014$ \\  
TW Nor & 10.786 & $421.2$ & 3.7 & 6.6 & 7.6 & $  5.653  \pm 0.022 $& $5.855  \pm 0.030$  & $0.27 \pm 0.10$ &  G15& $1.190 \pm 0.023$ \\ 
U Sgr & 6.745 & $1554.1 $& 2.7 & 5.4 & 6.0 &          & $3.565  \pm 0.028$    & $0.08 \pm 0.08$ & L11& $0.408 \pm 0.007$\\  
V Cen & 5.494 & $1336.5$& 2.2 & 6.1 & 6.5 & $ 4.293      \pm 0.014 $ & $4.258  \pm 0.028$    &  $0.04 \pm0.09$ & R08 & 0.265 $\pm$ 0.016 \\  
V Lac & 4.982 & $519.7$ & 4.7 & 6.6 & 8.1 & $  6.304 \pm 0.013 $ &                       & $ 0.06 \pm 0.06$ &   LL11 & $0.293 \pm 0.034$ \\ 
V0340 Nor & 11.289 & $ 513.2  $ & 1.4 & 6.5 & 6.6 & $ 5.293 \pm 0.005 $ & $5.313 \pm 0.029$& $0.07 \pm 0.07$ & G15& $0.312        \pm 0.050$ \\ 
V0367 Sct & 6.295 & $ 513.6 $& 2.7 & 6.7& 7.2 & $ 5.865  \pm 0.033 $ & $6.125 \pm 0.054$ & $ 0.05 \pm 0.08 $ &  G15 & \\ 
V0378 Cen$^{(*)}$ & 6.459 &  $517.8$& 3.7 & 6.5& 7.5 & $ 5.468 \pm 0.002 $ &             & $  0.08 \pm 0.06 $ &  LL11&  $ 0.374 \pm 0.049$\\ 
V0379 Cas$^{(*)}$ & 4.305 &$556.6$ &  2.2 & 6.5 & 6.9 & $ 5.876 \pm 0.005$ &             & $0.06 \pm 0.08$ &  L11& \\ 
V0438 Cyg & 11.211 & $561.9$ &  3.6 & 6.4 & 7.3 & $5.142         \pm 0.021$    &             & $ 0.33 \pm 0.06 $ & LL11 & \\  
VW Cru & 5.265 & $731.8$&  6.1 & 6.3 & 8.7 & $ 5.601 \pm 0.003$            &             & $0.19 \pm 0.06$ &  LL11 & 0.640       $\pm $ 0.046\\  
WX Pup & 8.933 & $344.7$&  4.6 & 6.5 & 8.0 & $ 6.323     \pm 0.004  $  & $6.326\pm0.016$ & $ -0.15 \pm 0.15 $ &  G14 &  $  0.306 \pm 0.018$  \\  
X Cru & 6.220 & $639.2$ &  2.5 & 6.0 & 6.5 &  $  5.677  \pm 0.008  $ &                   & $0.15 \pm 0.06$ &  LL11& $0.294       \pm 0.019$\\  
X Lac$^{(*)}$ & 5.443&  $519.7$& 4.7 & 6.6 & 8.1  & $  5.766    \pm 0.004 $ &            & $ 0.08\pm 0.06 $  &  LL11& \\  
X Pup & 25.972 &   $ 363.4 $& 3.7  & 5.1 & 6.4  & $ 4.089 \pm 0.345 $ &$5.023 \pm 0.012$ & $ 0.02 \pm  0.08$ &  G15 &  $0.396 \pm 0.015$\\  
X Vul & 6.320 & $ 879.5$& 3.2 & 6.0 & 6.8 & $ 4.727      \pm 0.007 $ & $4.860  \pm 0.031$    & $0.07\pm 0.08$ & L11&  $0.775 \pm 0.021$\\  
XZ Car &  16.652& $ 482.1$&  2.5 & 6.6 & 7.1  & $  5.269 \pm 0.033  $ & $5.169 \pm 0.010$& $0.19 \pm 0.06 $ &  LL11 & $        0.372 \pm 0.026$\\  
Y Sct & 10.341 & $ 503.4$ &  2.8 & 6.4 & 7.0 & $  5.121 \pm 0.171   $ &                  & $0.23 \pm 0.06 $ &   LL11 & $ 0.792 \pm 0.021$ \\  
               \bottomrule
\end{tabular}
 \tablefoot{Cluster average parallaxes include the corrections as described by L21.  Iron abundances  were rescaled by \citet{genovali2015alpha} 
 to the common solar abundance $A({\rm Fe})=7.50$  \citep{grevesse1998solar}. Color excess values E(B-V) are taken from \citet{Fernie} and scaled by a factor 0.94 \citep[cf.][]{groenewegen2018cepheid}. The symbol $^{(*)}$ denotes Cepheids pulsating in the first overtone mode. $^a$: Observations reported in the \hst\ system ($W_H$) are computed using Eq.\,\ref{eq:WH} and the individual passband data from \citet{2022arXiv220801045R}, that is, they do not contain the CRNL correction needed for comparison with the SN-host Cepheid sample.\\
    \textbf{References:} (LL11): \cite{luck2011distribution3}, (G14): \cite{genovali2014fine}, (L11): \cite{luck2011distribution2}, (G15): \cite{genovali2015alpha}, (R08): \cite{romaniello2008influence}, (L07): \cite{lemasle2007detailed}. }
    \label{tab:metal}
\end{table*}
\endgroup

\begin{figure}[ht!]
    \centering
   \includegraphics[scale=0.39]{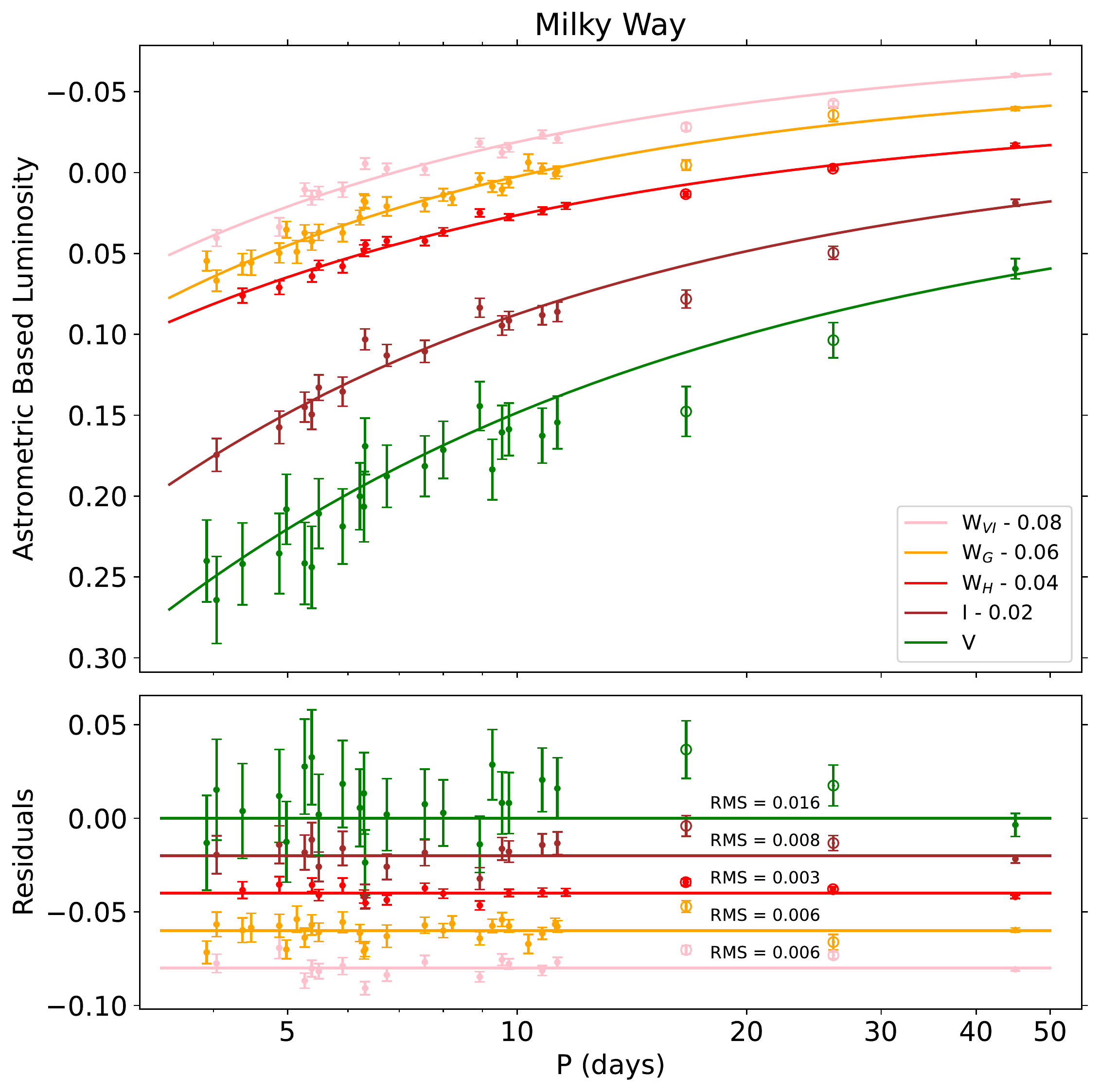}
    \caption{ABL for fundamental-mode Cepheids in the Gold sample using different photometric systems. Open circles indicate the two Cepheids of the Silver sample that are not part of the LL fits. The ABL values and the residuals were shifted by constant offsets as indicated in the legend to facilitate visual inspection. Cluster parallaxes were determined after applying the L21 parallax corrections.}
    \label{fig:ABL_optical}
\end{figure}

\subsection{Galactic LL and residual parallax offset for Cepheids \label{sec:combined}}

We calibrated the Milky Way LL and the residual parallax offset applicable to MW Cepheid parallaxes, $\Delta \varpi_{\rm Cep}$, using our Gold sample of cluster Cepheids. We note that the following exclusively considers MW Cepheid information and is thus independent of the LMC, which was merely used as a cross-check in Sect.\,\ref{sec:ZP}.

We fit the MW LL while simultaneously determining the residual parallax offset for Cepheids, $\Delta \varpi_{\rm Cep}$, using
\begin{align}
  \text{ABL} = 10^{M / 5} = 
  \begin{cases}
   \hfil    \varpi   10^{\frac{m -10 }{5}} &  \text{for clusters}, \\
    \hfil (\varpi+ \Delta \varpi_\text{Cep}) 10^{\frac{m -10 }{5}}
    & \text{for Cepheids},
\end{cases}
\end{align}
with $M = \alpha\cdot(\log{P} -1 ) + \beta = \alpha\cdot(\log{P} -1 ) + \delta + \gamma\cdot\mathrm{[Fe/H]}$. Both LL slope and zeropoint were used as free parameters, and $\Delta \varpi_{\rm Cl}=0$ as explained above.

\begin{figure*}[ht!]
    \centering
    \includegraphics[scale=0.5]{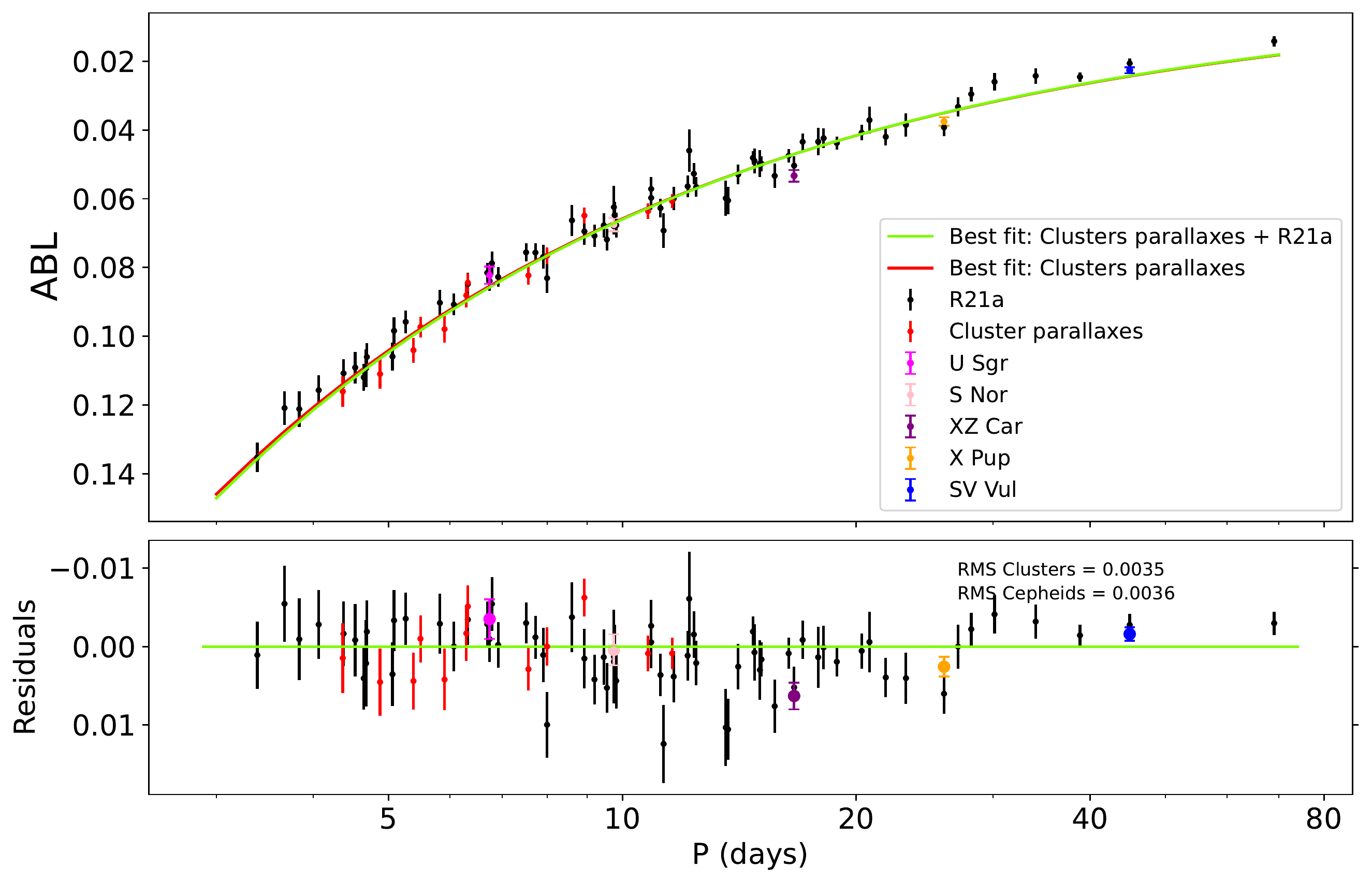}
\caption{ABL for $W_H$ based on \hst\ photometry for the joint sample of Gold cluster Cepheids (N = 15) and the Cepheids in the R21a sample (N = 67). Black error bars are derived using \gaia\ EDR3 parallaxes of Cepheids, and colored error bars are based on cluster parallaxes. Specific cases are colored individually to help identify Cepheids with cluster parallaxes discussed in the text.
U~Sgr, S~Nor, and SV~Vul appear twice in the plot  because  we use the Cepheid and cluster parallaxes to  estimate its ABL.   The Cepheids in the Silver sample  XZ~Car and X~Pup are not included in the fit. In the plot, the zeropoint offset of the Cepheids has been already applied.}
    \label{fig:JR}
\end{figure*}

We first performed this fit at the sample average iron abundance and then repeated the fit assuming a fixed value of $\gamma$ from the literature, specifically, $\gamma_{W_{H}} = -0.217 \pm 0.046 $ \citep{riess2021comprehensive} and $\gamma_{W_{G}} = -0.384 \pm 0.051$ \citep{breuval2022improved}. We used individual Cepheid iron abundances, not the sample average, to determine the zeropoint at solar metallicity, $\delta$.  Using  fixed literature slopes for $\gamma$ has the significant benefit of $\gamma$ being informed by a wider range of metallicities, while both the range of [Fe/H] in the MW sample and the correction to the solar value are small. Although we propagated the errors, this metallicity correction has virtually no effect on the final results due to the only slightly supersolar metallicity of MW Cepheids. Following common practice \citep[e.g.,][]{Kodric2018PHAT,riess2021comprehensive}, we applied a $2.7\sigma$ outlier rejection. This step removed $24$ of $249$ Cepheids for the \gaia-only sample, the vast majority of which are $> 3\sigma$ outliers. The ABL fit results are illustrated in Figures\,\ref{fig:JR} and \ref{fig:GDR3}.

\begin{figure*}[ht!]
    \centering
     \includegraphics[scale=0.5]{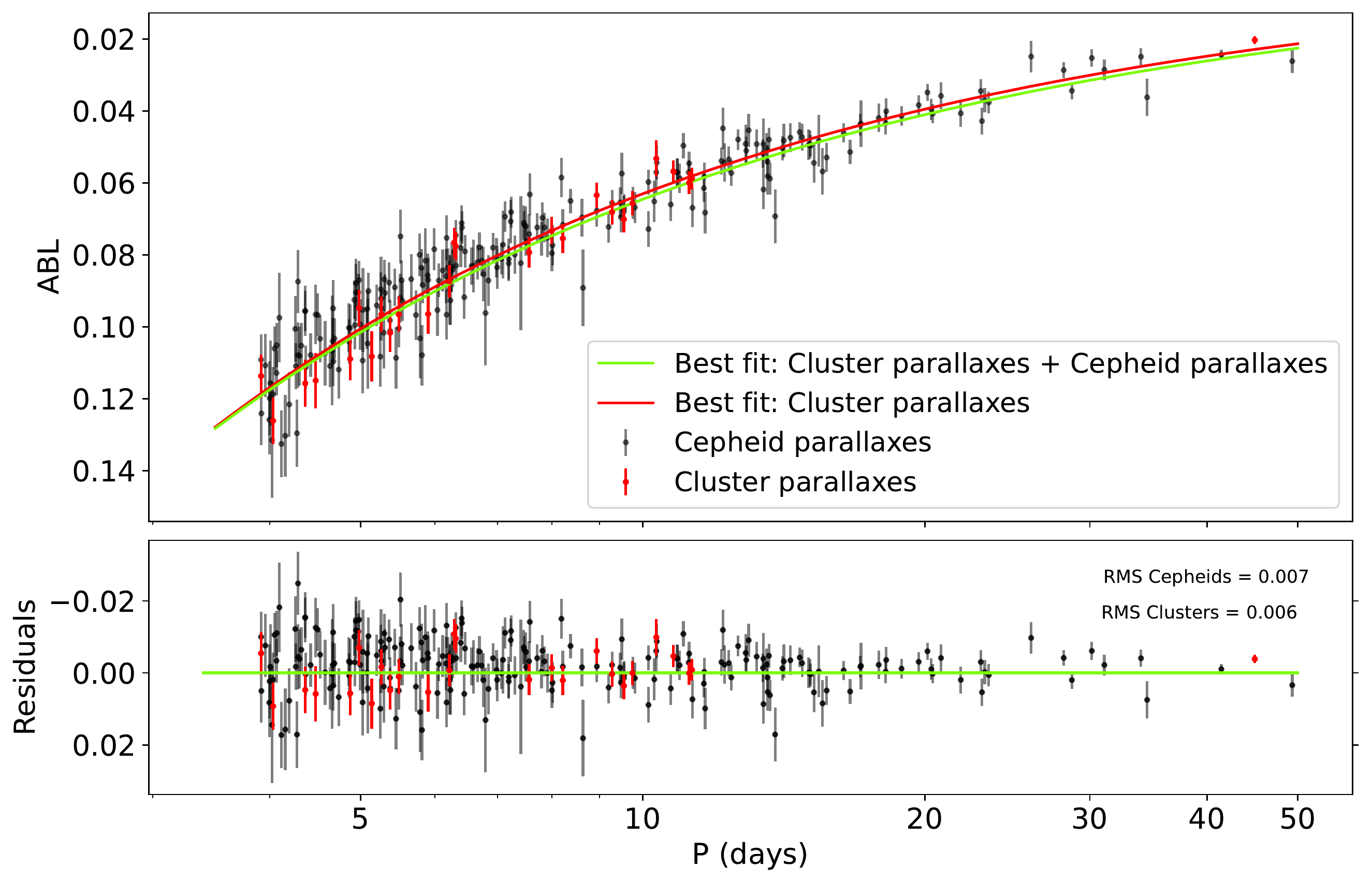}
\caption{ABL in the \gaia\ Wesenheit magnitude $W_G$ (cf. Figure~\ref{fig:JR} for sample details).}
    \label{fig:GDR3}
\end{figure*}

At the average sample metallicity, we find for the NIR Wesenheit LL ($W_H$)
\begin{align}
 \beta = -5.953 \pm 0.020, \quad \alpha = -3.406   \pm  0.052,\\
\Delta \varpi_\text{Cep} =  -18  \pm    5\, \mu \text{as}, \quad \chi^{2} =1.1, \\
\text{[Fe/H]} = 0.086 \ ,
\end{align}
which at solar metallicity becomes
\begin{align}
 \delta =  -5.930 \pm 0.020  , \quad \alpha =  -3.383  \pm 0.052 , \label{eq:WHdelta}\\
\Delta \varpi_\text{Cep} =  -17  \pm    5\, \mu \text{as}, \quad \gamma^{\text{fixed}}_{W_{H}} = -0.217 \pm 0.046,\\
 \chi^{2} =1.1, \quad \text{N}^{\text{MW}}_{\text{Cl}}= 15,  \quad \text{N}^{\text{MW}}_{\text{Cep}}= 67.   
\end{align}
We note that CRNL corrections ($\sim 0.05$\,mag) were applied to the apparent WFC3/IR F160W and NIR Wesenheit magnitudes to facilitate the comparison with Cepheids in supernova-host galaxies and simplify the comparison with the SH0ES distance ladder.

Both results establish a nonzero residual parallax offset for MW Cepheid parallaxes at $\gtrsim 3\sigma$ significance, and this result is fully consistent with the $-14 \pm 6\,\mu$as offset determined by \citet{riess2021cosmic}. This provides additional evidence that clusters and Cepheids require different residual parallax offsets. 

To directly compare our results to the value of $M_{H,1}^W$ determined as part of the SH0ES distance ladder \citep{riess2021comprehensive,2022arXiv220801045R}, we fixed the LL slope to the SH0ES baseline value and obtained
\begin{align}
 \delta =  -5.914 \pm 0.017  , \quad \alpha_{\text{fixed}} =  -3.299 \pm 0.015, \label{eq:WH2delta}\\
\Delta \varpi_\text{Cep} =  -13  \pm    5\, \mu \text{as}, \quad \gamma^{\text{fixed}}_{W_{H}} = -0.217 \pm 0.046,\\
 \chi^{2} =1.2.   
\end{align}
Our result for $\delta$ agrees to within $0.3\sigma$ with the value of $M_{H,1}^W$ determined by the SH0ES team via the two-parameter Gold sample fit in Table\,5 of \citet{2022arXiv220801045R}, where $M_{H,1}^W = -5.907 \pm 0.018$\,mag. Nevertheless, our approach to determine $\delta$ using the NIR Wesenheit function $W_H$ (Eq.\,\ref{eq:WHdelta}) differs from their approach in three important elements. First, we used a combined fit of Cepheid and cluster parallaxes to obtain an absolute calibration based exclusively on \gaia\ astrometry. Second, our clustering analysis in Sect.\,\ref{sec:clustering} was conducted entirely independently of \citet{2022arXiv220801045R}. Third, the samples of cluster member stars differ between our study and \citet{2022arXiv220801045R}, resulting in an average difference of $\sim 5\,\mu$as among cluster parallaxes. We therefore consider our result an important cross-check based on mostly independent astrometric information.

For the corresponding \gaia\ Wesenheit function ($W_G$) at sample average metallicity, we obtain
\begin{align}\label{eq:WGavgZ}
 \beta = -6.051  \pm 0.020 , \quad \alpha = -3.303 \pm 0.049,\\ 
\Delta \varpi_\text{Cep} =  -22  \pm    3\, \mu \text{as}, \quad \chi^{2} =1.2, \\
\text{[Fe/H]} = 0.069 \ , 
\end{align}
and, after correcting to solar metallicity using the individual Cepheid iron abundances,
\begin{align}
 \delta=  -6.004   \pm 0.019, \quad \alpha = -3.242 \pm 0.047,\\
\Delta \varpi_\text{Cep} =  -19  \pm    3\, \mu \text{as}, \quad \gamma^{\text{fixed}}_{W_{G}} = -0.384 \pm 0.051,\label{eq:offset_gaia} \\
  \chi^{2} =1.1,\quad  \text{N}^{\text{MW}}_{\text{Cl}}= 26,  \quad \text{N}^{\text{MW}}_{\text{Cep}}= 225.
\end{align}

We thus find $1\sigma$ agreement for $\Delta \varpi_{\rm Cep}$ regardless of whether \hst\ or \gaia\ photometry is used, and using different, albeit not independent, sets of Cepheids and cluster parallaxes. In particular, we note the improved precision on $\Delta \varpi_{\rm Cep}$ determined using \gaia\ photometry, for which we obtain a $6\sigma$ detection that is consistent with the value determined using the independent \hst\ photometry. We further note that metallicity corrections do not challenge the accuracy of our determination of $\Delta \varpi_{\rm Cep}$.  To illustrate our results in a more conventional LL form, we plot the absolute Wesenheit magnitudes as a function of $\log{P}$ in Figure~\ref{fig:Leavitt}. 
\begin{figure*}[ht!]
    \centering
     \includegraphics[scale =0.5]{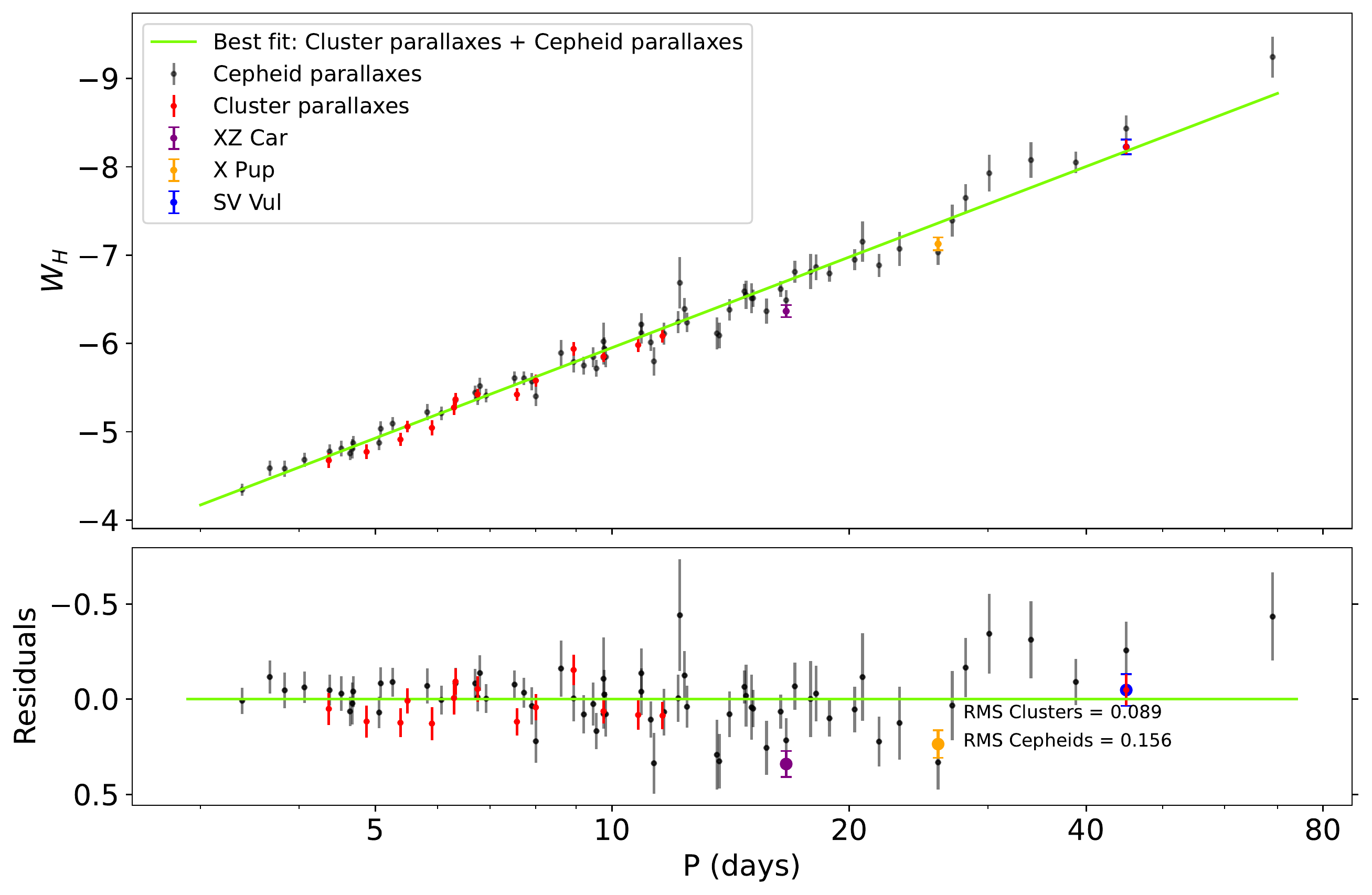}
    \includegraphics[scale =0.5]{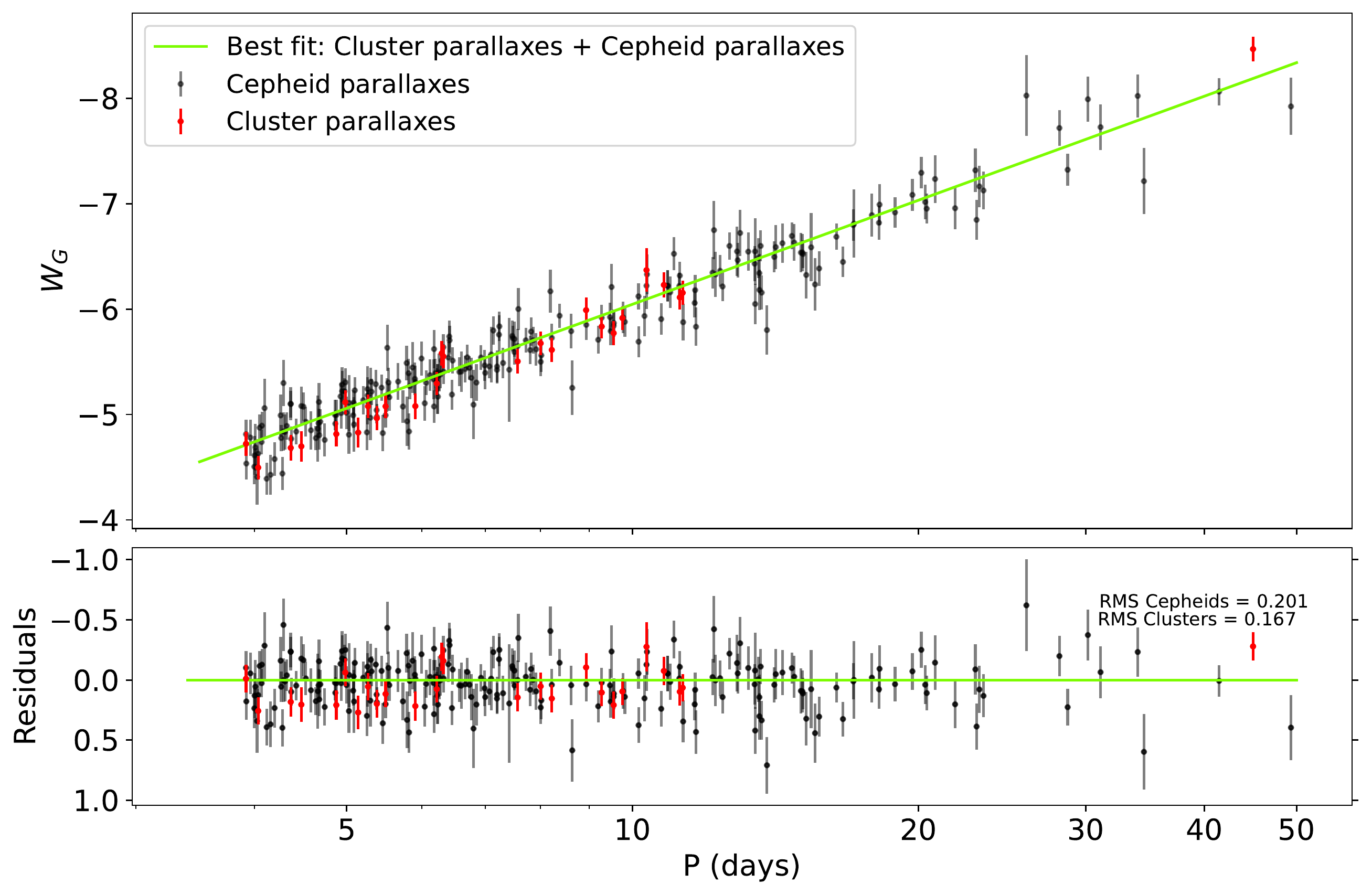}
    \caption{LL in the $H$ and $G$ Wesenheit bands. Given the high precision of the Cepheid parallaxes, their individual distances  were calculated as $1/\varpi$.  The plots are shown for illustration purposes, and they were not used to fit the data. }
    \label{fig:Leavitt}
\end{figure*}

\begingroup
\setlength{\tabcolsep}{6.pt} 
\begin{table*}[]
    \centering
    \caption{Combined fits to cluster Cepheids and (field) MW Cepheids. }
    \begin{tabular}{@{}lcccccc@{}}\toprule
         Filter & $\alpha$ & $\beta$ & $\Delta \varpi_{\rm Cep}$ & $\langle\mathrm{[Fe/H]}\rangle$ & N$_{\mathrm{Cep}}$ & N$_{\mathrm{cl}}$ \\
          & (mag/$\log{P}$) & (mag) & ($\mu$as) & & &\\\midrule
         $W_H^{a}$ & $ -3.412 \pm 0.053 $ & $ -6.003 \pm 0.020$ & $ -19 \pm 5 $ & 0.086 & 67& 15 \\
         $W_G$ & $-3.303 \pm 0.049$ & $-6.051 \pm 0.020$ & $-22 \pm 3$ & 0.069 & 225& 26\\\midrule
         $Bp$  & $-2.513 \pm 0.080 $ & $ -4.225 \pm 0.036$ & $-21 \pm 6$ & 0.069& 243& 23 \\ 
         $V$   & $-2.553 \pm 0.071$ & $-4.377 \pm 0.033 $ &$-22 \pm 5$ & 0.069& 246 & 22\\
         $G$   & $-2.751 \pm 0.077 $ & $-4.612 \pm 0.035$ & $-22 \pm 5$ & 0.068& 238 & 23\\
         $Rp$  & $-2.804 \pm 0.058$ & $-4.984 \pm 0.025$ & $-19\pm 4 $& 0.068& 234& 23\\
         $F160W^{a}$  & $-3.353 \pm 0.060$ & $-5.729 \pm 0.023$ & $-23\pm 6$ & 0.088& 67    & 15    \\\midrule
         $W_H^{b}$ & $-3.406 \pm 0.052$ & $-5.953 \pm 0.020$ & $-18 \pm 5$ & 0.086& 67& 15\\
         $F160W^{b}$  & $-3.346 \pm 0.060$ & $-5.679 \pm 0.023$ & $-22\pm 6$ & 0.088& 67& 15\\
         \bottomrule
    \end{tabular}
    \tablefoot{$^{a}$ Does not include the CRNL correction. $^{b}$ Includes the CRNL correction (mean  $0.010$\,mag) to facilitate comparison with extragalactic Cepheid samples  \citep{2022arXiv220801045R}. }
    \label{tab:MWfits}
\end{table*}
\endgroup

We further applied the same approach for Johnson $V-$band, \gaia\ $G$, $Bp$, and $Rp$, and \hst\ $F160W$ photometry. The results are listed in Table\,\ref{tab:MWfits}. In particular, we note that the value of $\Delta \varpi_{\rm Cep}$ is consistent within less than $1 \sigma$ for all nine rows in Table\,\ref{tab:MWfits}. Figure \ref{fig:alpha-beta}  illustrates the results for individual photometric passbands together with linear fits of the LL parameters as a function of the inverse of the effective central wavelength $\lambda$ of each filter. The average iron abundances of the samples  differ by $< 0.02$\,dex, and we thus expect  a difference of  $\sim$ 0.02 dex $\cdot$ 0.2 mag/dex  = 0.004\,mag at most between the values of $\beta$ evaluated at the lower and upper metallicity of our sample. This difference is well contained within the uncertainties. Fitting the wavelength dependence of $\alpha$ and $\beta$ as a function of inverse wavelength, we determine the following dependence of LL slope and zeropoint on central wavelength:

\begin{align}
\alpha &= (-3.769 \pm 0.083) + (0.683 \pm 0.059)/\lambda \label{eq:alpha}\\
\beta &= (-6.526 \pm 0.056) + (1.208 \pm 0.041)/\lambda\label{eq:beta} \ .
\end{align}

\begin{figure*}[ht!]
    \centering
     \includegraphics[width=0.99\textwidth]{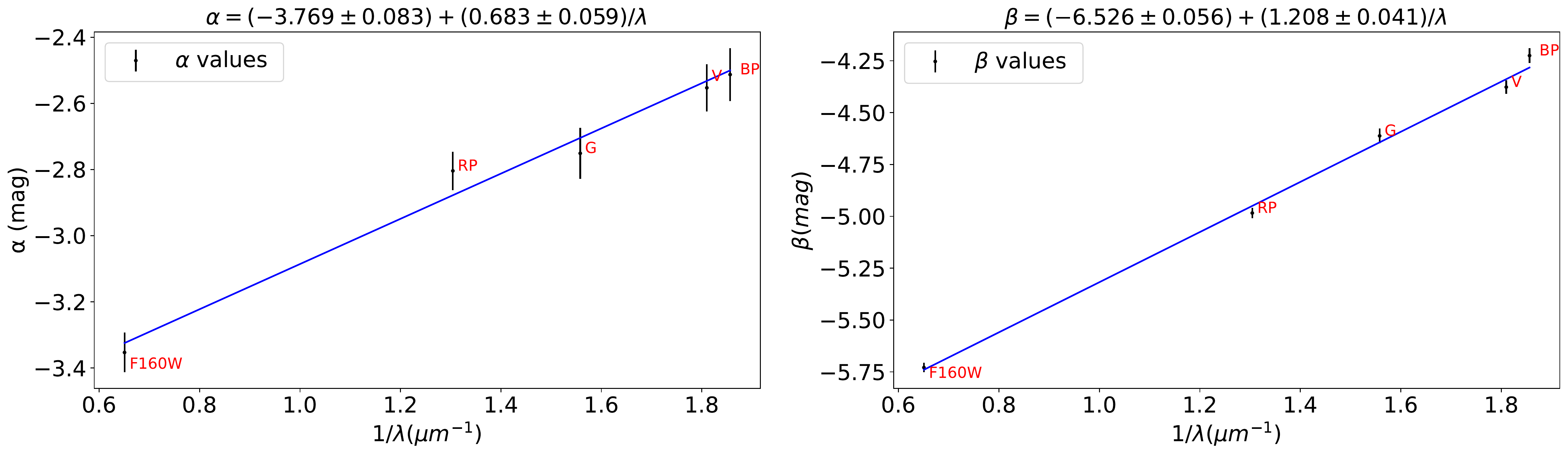}
    \caption{Linear fit of the LL parameters as a function of the inverse of the effective wavelength in different photometric filters.}
    \label{fig:alpha-beta}
\end{figure*}

\section{Discussion}\label{sec:Discussion}
\subsection{Using Silver sample Cepheids for LL calibration\label{sec:samples}}

Our criteria placed two long-period Cepheids with uncertain cluster membership, X~Pup and XZ~Car, in the Silver sample, which we conservatively did not use for LL calibration. As explained in Sec.\,\ref{sec:silver}, both stars featured low membership likelihoods due primarily to mismatching kinematic information. However, closer inspection suggested that X~Pup is  possibly a true cluster Cepheid that can be used for LL calibration (cf. Sec.\,\ref{sec:silver}). We determine the impact of including these stars in our analysis below.

Including X~Pup and XZ~Car in the cross-check of involving the LMC (Sect.\,\ref{sec:ZP}) would not significantly affect the results. For $W_G$ we find $\Delta \varpi_{\rm Cl} = -8 \pm  17 \,\mu$as, $\gamma = -0.418 \pm 0.150$ mag/dex, and for $W_H$, we obtain $\Delta \varpi_{\rm Cl} = -7 \pm  16 \,\mu$as, $\gamma = -0.205 \pm 0.148$ mag/dex. All these values agree to within much less than one standard deviation with those obtained using only the Gold sample of Cepheids.

Including XZ~Car in the combined LL fit in Sec.\,\ref{sec:combined} has no impact because it is a $3.5\sigma$ LL outlier that would be rejected by the $\sigma-$clipping procedure. Including X~Pup in the fit does not significantly affect the LL calibration ($\alpha=-3.313\pm 0.049$, $\beta = -6.051 \pm 0.020$, $\Delta \varpi_{\rm Cep}=-21 \pm 3$ all agree to much better than $1\sigma$ with results in Eq.\,\ref{eq:WGavgZ}) and marginally increases the reduced $\chi^2$ by $0.008$. Furthermore, X~Pup has not been identified as an LL outlier by \citet{2022arXiv220801045R} in the NIR Wesenheit formulation.

\subsection{Fraction of Cepheids in clusters within $2$\,kpc \label{sec:fcc}} 

The fraction of Cepheids residing in clusters is of interest for understanding clustered star formation \citep{dinnbier2022dynamical} and the extragalactic distance scale \citep{anderson2018cepheid}, among other things. Using our Gold sample of cluster Cepheids and data from the recent \gaia\ DR3, we updated previous estimates of this fraction, $f_\mathrm{CC,2kpc} = N_{\mathrm{Cl,2kpc}} / N_{\mathrm{Cep,2kpc}}$. Assuming that all Cepheid-hosting clusters within $2$\,kpc could be identified by our method, we have $N_{\mathrm{Cl,2kpc}}=22$ (Gold sample), which includes $11$ coronal members separated by projected distances of $8-25$\,pc from their host cluster centers.  

We estimated the total number of Cepheids within $2$\,kpc, $N_{\mathrm{Cep,2kpc}}$ using the photometric parallaxes obtained with our $W_G$ LL calibration for all stars classified as \texttt{DCEP} in \gaia\ DR3 table \texttt{gaiadr3.vari\_cepheid}. This yields $180^{+32}_{-38}$ fundamental mode Cepheids as well as $70^{+9}_{-16}$ first-overtone or multimode Cepheids, where overtone periods were fundamentalized using the period ratios determined by \citet[assuming a mean metallicity $\langle \mathrm{[Fe/H]} \rangle = 0.032$]{kovtyukh2016chemical}. For multimode Cepheids, either the fundamental or first-overtone period was used to compute the distance. We also sought to estimate $N_{\mathrm{Cep,2kpc}}$ using distances provided by the parameter \texttt{distance\_gspphot} in \gaia\ DR3 table \texttt{gaiadr3.gaia\_source} as well as \gaia\ parallaxes (including the residual offset determined in Eq.\,\ref{eq:offset_gaia}). However, this reduced the size of Cepheid samples by approximately $20\%$ due to limited data availability. We therefore considered the estimation based on photometric distances our baseline result due to greater completeness.
The results are tabulated in Table\,\ref{tab:NCepheids}, where asymmetric uncertainties reflect the range of stars defined by the $1\sigma$ distance or parallax uncertainties. 

We thus estimate $f_{\mathrm{CC,2kpc}} = 0.088^{+0.029}_{-0.019}$, where the uncertainties provided denote the full range of possibilities. We further find a slightly higher fraction of fundamental mode Cepheids in clusters, with $f_{\mathrm{CC,2kpc,FM}} = 0.089^{+0.030}_{-0.018}$ and $f_{\mathrm{CC,2kpc,FO}}=0.081^{+0.026}_{-0.023}$, assuming the OGLE classification of cluster Cepheids \citep{pietrukowicz2021classical}. If the pulsation modes assigned in \gaia\ DR3 \citep{gaiadr3.cepheid} were used instead, the difference would be slightly larger, with $f_{\mathrm{CC,2kpc,FM}} = 0.100^{+0.033}_{-0.020}$ and $f_{\mathrm{CC,2kpc,FO}}=0.065^{+0.021}_{-0.022}$. This difference could be explained by the dependence of $f_{\rm CC}$ on age due to clusters dissolving into the field over time combined with the tendency of overtone Cepheids to originate from older lower-mass stars than fundamental-mode Cepheids, which can be rather young.

We note that a few bright Cepheids, such as Polaris and the cluster Cepheid U~Sgr, are not included in the vari\_cepheid table. However, their absence does not change the overall result. Our new estimate supersedes our previous slightly lower estimate of $f_{CC,2kpc} = 15/217 = 6.9\%$ reported in \citet{dinnbier2022dynamical} due to improvements in our membership determination and the input data from \gaia\ DR3.

Figure\,\ref{fig:fcc} illustrates the fraction of Cepheids residing in clusters within 2\,kpc of the Sun as a function of age. Cepheid ages were computed using period-age relations for fundamental and first-overtone Cepheids (\citet{anderson2016effect}). We confirmed that ages based on periods of overtone Cepheids matched ages computed using period-age relations for fundamental-mode Cepheids after fundamentalizing the pulsation periods of first-overtone Cepheids using period ratios of Milky Way double-mode Cepheids \citep{kovtyukh2016chemical}. Figure\,\ref{fig:fcc} thus illustrates the dispersal of Cepheid host clusters over time, an effect previously reported by \citet{anderson2018cepheid} and also seen in dynamical NBODY simulations \citep{dinnbier2022dynamical}. We caution that young ages are rather poorly sampled within $2$\,kpc of the Sun due to the low volumetric rate of long-period Cepheids. At ages above 132\,Myr, no cluster Cepheids are found within $2\,$kpc of the Sun.

\begin{figure}[ht!]
    \centering
     \includegraphics[scale = 1.03]{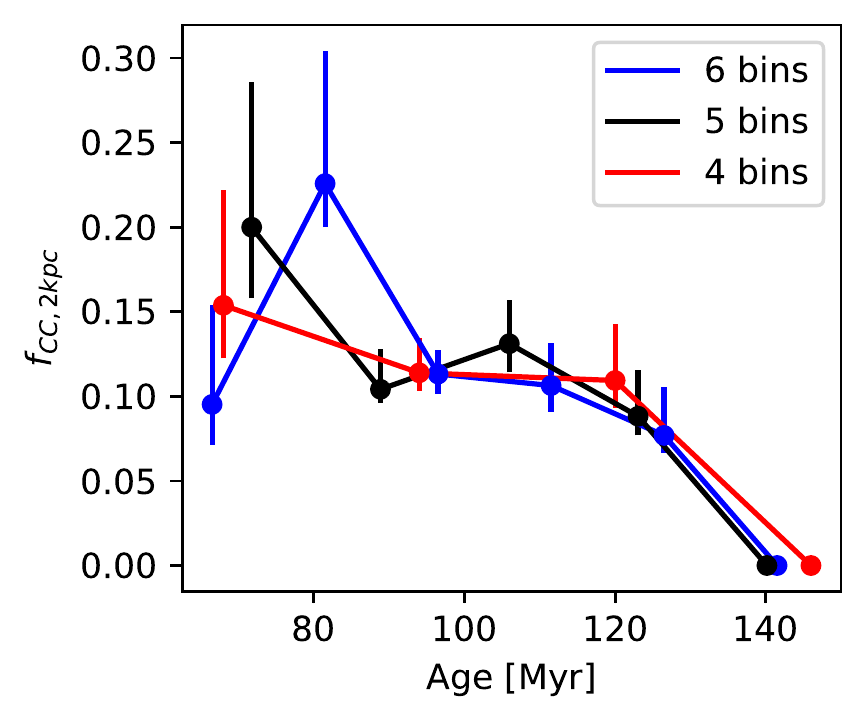}
    \caption{Clustered Cepheid fraction as a function of Cepheid age estimated using the period-age relations for solar metallicity \citep{anderson2016effect}. The size of the error bars illustrates the full range of possible fractions. Different numbers of bins were used to illustrate the dependence on binning. Young long-period Cepheids are rare within 2\,kpc of the Sun, increasing the scatter at ages below $80$\,Myr. No Cepheids older than $132$\,Myr are found in clusters within 2\,kpc of the Sun.}
    \label{fig:fcc}
\end{figure}

\begingroup
\setlength{\tabcolsep}{7.pt} 
\renewcommand{\arraystretch}{1.5}
\begin{table*}[]
    \centering
    \caption{Number of Cepheids within 2 kpc of the Sun. }
    \begin{tabular}{cccccc}\toprule
         Sample & Method & Fundamental mode & Overtone & Multi mode & Total \\\midrule
         Field & LL $W_G$ & $180^{+32}_{-38}$ & $62^{+9}_{-15}$ &  $8^{+0}_{-1} $ & $250^{+41}_{-54}$ \\ 
         Field & distance\_gspphot  &   $139^{+12}_{-13}$& $55^{+7}_{-9}$ & $7^{+0}_{-1}$& $201^{+19}_{-23}$\\
         Field & Parallax   &   $130^{+10}_{-13}$& $61^{+6}_{-13}$ & $7^{+1}_{-0}$ & $198^{+17}_{-26}$ \\
         Clusters & LL $W_G$, OGLE classifiers &   $16^{+1}_{-1}$ & $5^{+0}_{-1}$ & $1^{+0}_{-0}$& $22^{+1}_{-2}$ \\
         Clusters & LL $W_G$, Gaia classifiers &   $18^{+1}_{-1}$ & $4^{+0}_{-1}$ & $0^{+0}_{-0}$& $22^{+1}_{-2}$ \\
         \bottomrule
    \end{tabular}
    \tablefoot{The upper and lower indexes are an estimate of the maximum and minimum number of Cepheids, they are not standard errors, and  for this reason, they are not are not added in quadrature.}
    \label{tab:NCepheids}
\end{table*}
\endgroup

\subsection{Expected improvements\label{sec:improvements}}

Astrometric uncertainties tend to increase with distance, complicating the identification of distant open clusters. For \gaia\ EDR3, the number of false cluster detections  at distances  greater than $3$ kpc increases rapidly, so that significant work is required to ascertain the veracity of the recovered cluster candidates.  However, upcoming \gaia\ data releases will improve the ability to correctly identify clusters at large distances, which can be expected to result in much improved cluster Cepheid samples with \gaia\ DR4 and beyond. Whereas \gaia\ EDR3 was based on 34 months of observations, the DR4  astrometric solution of \textit{Gaia}  will be based on approximately 66 months of observations, and the \gaia\ Collaboration expects improvements in proper motion proportional to $t^{-3/2}$ and in parallax proportional to $t^{-1/2}$. Hence, DR4 proper motion uncertainties may be about $0.35$ times their DR3 uncertainties, whereas DR4 parallax uncertainties could be approximately $0.70$ times those reported in DR3. As Eq.\,\ref{eq:1} illustrates (cf. also footnote\,\ref{foot:pm}), the ability of detecting clusters against the background depends on distance and proper motion uncertainties. However, it is unlikely that the full gain in proper motion precision will directly map to a greater volume limit for detecting clusters because parallax errors improve less rapidly. To obtain a rough estimate of future improvements, we therefore considered a mean improvement by a factor of approximately 2 (counting parallax and both proper motion directions separately), which would double the distance within which cluster Cepheids can be detected. Based on their location in the Galactic plane, the number of clusters increases proportional to $d^2$, resulting in a potential quadrupling of cluster-hosting Cepheids with DR4, and thus, in a potential improvement of a factor of $2$  for the LL calibration. Since most long-period Cepheids are located at distances beyond $2\,$kpc, this will be particularly useful to increase the number of these high-priority targets.

Calibrating the cosmic distance ladder to within $1\%$ requires parallaxes of Cepheids measured to an accuracy of $ \sim 5 \,\mu$as \citep{riess2021cosmic}. At present, cluster Cepheids appear to be the most viable route to this goal. However, the angular covariance of the (E)DR3 parallaxes currently still sets an error floor  of $\sim 7\,\mu$as and is therefore in urgent need of further improvement. It is very  noteworthy that cluster members apparently do not require residual parallax offset corrections, since solving for this offset has thus far limited the power of \gaia\ parallaxes for measuring \Ho\ \citep[e.g.,][]{riess2018standards,riess2021cosmic}. Additionally, new \hst\  observations of cluster Cepheids will be crucial to avoid uncertainties related to photometric transformations from the ground to the \hst\ system.  In summary, identifying new cluster Cepheids and measuring their photometry using \hst\ will provide the most accurate basis for calibrating the distance ladder for a $1\%$ \Ho\ measurement. We are optimistic that future \gaia\ data releases will continue to improve the error floor set by angular covariance and that other mitigation strategies can be identified to leverage the \  power of \textit{Gaia} for the extragalactic distance scale and cosmology.

\section{Conclusions}\label{sec:Conclusions}
We carried out a systematic search for MW cluster Cepheids using Gaia EDR3 and DR3 data. The improved  proper motion precision of EDR3 over DR2 allowed us to obtain a more detailed and accurate view of cluster membership for previously discussed cluster Cepheids. Since our method requires no advance knowledge of clusters being present in the vicinity of Cepheids, we a) determined cluster astrometry without the need for prior literature search on the host clusters, and b) avoided confusion of cluster identification in case of complex sky areas featuring multiple clusters.
We thus established a Gold sample of 34 Cepheids residing in 28 distinct MW open clusters. They include the three new bona fide cluster Cepheids  ST Tau, V0378 Cen, and GH Lup. Additionally, we corrected the host cluster identification for three Cepheids previously discussed in the literature, namely SX Vel, IQ Nor, and VW Cru. We find  SV~Vul to be a bona fide cluster Cepheid that falls squarely on the Galactic LL. We find three Silver sample cluster Cepheid candidates of interest, of which X~Pup is a likely cluster Cepheid, whereas the XZ~Car cluster membership is tentatively excluded by kinematic constraints and the AP~Vel parallax narrowly contradicts membership in Ruprecht\,65. Additional combinations of possible interest are included in a Bronze sample.

Using photometric distances of Cepheids in the Gold sample and the concatenated list of Cepheids from \citet{pietrukowicz2021classical} and \citet{gaiadr3.cepheid}, we estimate the fraction of clustered Cepheids within $2$\,kpc to be in the range of  $f_{\mathrm{CC,2kpc}} = N_{\mathrm{Cl,2kpc}} / N_{\mathrm{Cep,2kpc}} = 0.088^{+0.029}_{-0.019}$. We find a slightly larger fraction for Cepheids pulsating in the fundamental mode compared to the first overtone, which may be related to the dependence of $f_{\rm CC}$ on age and cluster dispersal timescales.

Cluster parallaxes are superior for LL calibration compared to individual Cepheid parallaxes because cluster member stars combine several benefits, including a) greater statistical precision, b) better systematics in a fainter magnitude range that does not require special processing, c) the absence of high-amplitude variability, and d) greater consistency in brightness and color with LMC stars and quasars used to determine the EDR3 parallax systematics (L21). The uncertainty of average cluster parallaxes is currently dominated by angular covariance, which  limits average parallax uncertainties to $\gtrsim 7\,\mu$as, although the statistical uncertainty can be as low as $1.4\,\mu$as.

We identified the magnitude and color ranges of $12.5 < G < 17$\,mag and $0.23 < Bp - Rp < 2.75$ as a sweet spot for determining average cluster parallaxes. Previous studies \citep[e.g,.][]{flynn2022clusters,apellaniz2022clustersclouds} found that parallaxes of cluster member stars in this magnitude range are adequately corrected by the L21 recipes, and we cross-checked this result using Cepheids in the LMC, taking the metallicity difference between MW and LMC Cepheids into account. Using the LL metallicity slope $\gamma$ yielded negative values in six individual photometric passbands and three reddening-free Wesenheit magnitudes, confirming recent results by \citet{breuval2022improved}. Allowing for a nonzero offset for cluster parallaxes yields a weighted average of $\Delta \varpi_{\rm Cl} = -4 \pm 6\,\mu$as, with each individual offset consistent with $0$ to within $1\sigma$. Hence, we confirm that cluster parallaxes determined using member stars in this magnitude and color range require no further correction of residual parallax offsets beyond the L21 corrections. We stress that the LMC was used only for comparison and does not otherwise enter the results of this study.

Setting $\Delta \varpi_{\rm cl} = 0$, we calibrated the Galactic Cepheid LL in the several passbands and reddening-free Wesenheit magnitudes while simultaneously solving for a residual parallax offset of \gaia\ parallaxes of Cepheids, $\Delta \varpi_{\rm Cep}$. In particular, we calibrated the absolute luminosity scale of $10$\,d fundamental-mode Cepheids at solar metallicity to a precision of $0.94\%$ using NIR \hst\ Wesenheit magnitudes and to a precision of $0.87\%$ using optical \gaia\ Wesenheit magnitudes. The LL slope and metallicity effect from the SH0ES analysis provide the most direct comparison of our results of relevance for the Hubble constant and reveals excellent ($0.3\sigma$) agreement with the recent results by \citet{2022arXiv220801045R}. Using NIR \hst\  and optical \gaia\ Wesenheit magnitudes, we obtained $\Delta \varpi_{\rm Cep} = -17\pm5$ and $-19 \pm 3\,\mu$as, respectively. This $7\sigma$ measurement of the residual parallax offset for Cepheids is the most precise to date and provides strong independent confirmation of the Cepheid parallax offset of $-14 \pm 6\,\mu$as measured by the SH0ES team.

Cluster Cepheids can play a crucial role for the measurement of $H_0$ by providing an accurate absolute trigonometric scale based on \gaia\ astrometry without the need to solve for further offsets while determining the Hubble constant. Future developments, such as improved proper motion membership constraints for cluster detection through the longer astrometric baselines of \gaia\  in future data releases, improved corrections of the \gaia\  parallax systematics and angular covariance, and high-quality photometry of MW Cepheids in and out of clusters will particularly improve the base calibration of the distance scale toward a $1\%$ Hubble constant measurement.

\begin{acknowledgements}
We thank the anonymous referee for comments that allowed to improve the quality of the manuscript. We are thankful for useful discussions with Adam Riess and Stefano Casertano as well as their comments on an earlier version of the manuscript. We further acknowledge useful discussions in the framework of the ISSI International Team project \#490 meeting in Bern in July 2022.\\

MC \& RIA acknowledge support from the European Research Council (ERC) under the European Union's Horizon 2020 research and innovation programme (Grant Agreement No. 947660). RIA further acknowledges support through a Swiss National Science Foundation Eccellenza Professorial Fellowship (award PCEFP2\_194638). 

This work has made use of data from the European Space Agency (ESA) mission
{\it Gaia} (\url{https://www.cosmos.esa.int/gaia}), processed by the {\it Gaia}
Data Processing and Analysis Consortium (DPAC,
\url{https://www.cosmos.esa.int/web/gaia/dpac/consortium}). Funding for the DPAC
has been provided by national institutions, in particular the institutions
participating in the {\it Gaia} Multilateral Agreement. 

This research has made use of NASA's Astrophysics Data System; the SIMBAD database and the VizieR catalog access tool\footnote{\url{http://cdsweb.u-strasbg.fr/}} provided by CDS, Strasbourg; Astropy\footnote{\url{http://www.astropy.org}}, a community-developed core Python package for Astronomy \citep{astropy:2013, astropy:2018}; TOPCAT\footnote{\url{http://www.star.bristol.ac.uk/~mbt/topcat/}} \citep{2005ASPC..347...29T}.

\end{acknowledgements}

\begin{appendix}

\section{Detected host clusters\label{app:PMdispersion}}
Figure\,\ref{{fig:PMdispersion}} shows the estimated dispersion in proper motion for the detected clusters in this study.  Table\,\ref{tab:GaiaSample2} provides the Gaia~EDR3 source~ids for all cluster members, as well as the parallax corrections that were used to compute the residual offset for MW Cepheids in Sect.\,\ref{sec:combined}.  

\begin{figure*}[ht!]
    \centering
    \includegraphics[width=0.99\textwidth]{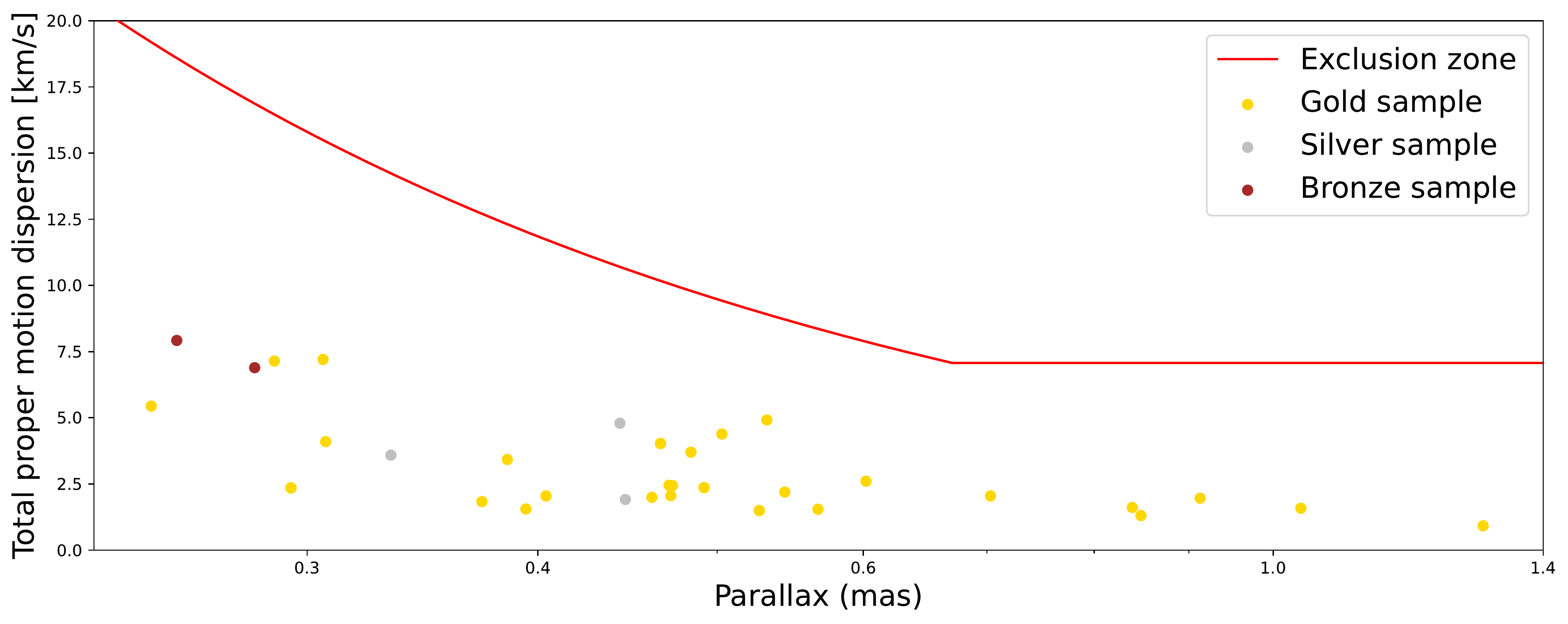}
    \caption{Proper motion dispersion of the detected clusters in the Gold, Silver, and Bronze samples. }
    \label{{fig:PMdispersion}}
\end{figure*}

\begin{table*}[!htp]
    \centering
    \caption{Cluster members of the Gold, Silver, and Bronze samples.}
    \begin{tabular}{@{}llll}\toprule
         Cluster & Gaia EDR3 source id  & $\varpi_{corr}$ & $corr$  \\
          &   & (mas) & (mas)  \\ \midrule
          Czernik 41 & 2026716726778925568 & 0.3398 & -0.0148    \\
          Czernik 41 & 2026716692420403328 & 0.4367  & -0.0408    \\
          Czernik 41 & 2026716731097672704 & 0.3948 & -0.0470   \\
          \ldots & \ldots & \ldots & \ldots \\
         \bottomrule
    \end{tabular}
    \tablefoot{The complete version of this table is available at the CDS. $\varpi_{\rm corr}$ is the  parallax corrected applying the L21 offset, and $\rm corr$ is the corresponding value of the correction.}
    \label{tab:GaiaSample2}
\end{table*}

\end{appendix}

\bibliographystyle{aa}
\bibliography{refs}
\end{document}